\documentclass[pra,aps,twocolumn,superscriptaddress]{revtex4-1}

\usepackage{graphicx}  
\usepackage{epstopdf}
\usepackage{dcolumn}   
\usepackage{bm}        
\usepackage{amssymb}   
\usepackage{amsmath}   
\usepackage{amsthm}
\usepackage{chemarrow}
\usepackage{color}
\usepackage{mathrsfs}
\usepackage{float}
\usepackage{bbold}
\usepackage{bbm}
\usepackage{ulem}
\usepackage{braket}
\usepackage[a4paper,colorlinks=true,
linkcolor=blue,citecolor=blue,
pdfauthor={ },
pdftitle={ },
pdfsubject={ },
pdfkeywords={ }]{hyperref}

\theoremstyle{plain}
\newtheorem*{theorem*}{Theorem}

\begin{document}


\title{Zero- to ultralow-field nuclear magnetic resonance and its applications}


\author{Min Jiang}
\affiliation{
Hefei National Laboratory for Physical Sciences at the Microscale and Department of Modern Physics, University of Science and Technology of China, Hefei 230026, China}
\affiliation{
CAS Key Laboratory of Microscale Magnetic Resonance, University of Science and Technology of China, Hefei 230026, China}
\affiliation{
Synergetic Innovation Center of Quantum Information and Quantum Physics, University of Science and Technology of China, Hefei 230026, China}

\author{Ji Bian}
\affiliation{
Hefei National Laboratory for Physical Sciences at the Microscale and Department of Modern Physics, University of Science and Technology of China, Hefei 230026, China}
\affiliation{
CAS Key Laboratory of Microscale Magnetic Resonance, University of Science and Technology of China, Hefei 230026, China}
\affiliation{
Synergetic Innovation Center of Quantum Information and Quantum Physics, University of Science and Technology of China, Hefei 230026, China}

\author{Qing Li}
\affiliation{
Hefei National Laboratory for Physical Sciences at the Microscale and Department of Modern Physics, University of Science and Technology of China, Hefei 230026, China}
\affiliation{
CAS Key Laboratory of Microscale Magnetic Resonance, University of Science and Technology of China, Hefei 230026, China}
\affiliation{
Synergetic Innovation Center of Quantum Information and Quantum Physics, University of Science and Technology of China, Hefei 230026, China}

\author{Ze Wu}
\affiliation{
Hefei National Laboratory for Physical Sciences at the Microscale and Department of Modern Physics, University of Science and Technology of China, Hefei 230026, China}
\affiliation{
CAS Key Laboratory of Microscale Magnetic Resonance, University of Science and Technology of China, Hefei 230026, China}
\affiliation{
Synergetic Innovation Center of Quantum Information and Quantum Physics, University of Science and Technology of China, Hefei 230026, China}

\author{Haowen Su}
\affiliation{
Hefei National Laboratory for Physical Sciences at the Microscale and Department of Modern Physics, University of Science and Technology of China, Hefei 230026, China}
\affiliation{
CAS Key Laboratory of Microscale Magnetic Resonance, University of Science and Technology of China, Hefei 230026, China}
\affiliation{
Synergetic Innovation Center of Quantum Information and Quantum Physics, University of Science and Technology of China, Hefei 230026, China}

\author{Minxiang Xu}
\affiliation{
Hefei National Laboratory for Physical Sciences at the Microscale and Department of Modern Physics, University of Science and Technology of China, Hefei 230026, China}
\affiliation{
CAS Key Laboratory of Microscale Magnetic Resonance, University of Science and Technology of China, Hefei 230026, China}
\affiliation{
Synergetic Innovation Center of Quantum Information and Quantum Physics, University of Science and Technology of China, Hefei 230026, China}

\author{Yuanhong Wang}
\affiliation{
Hefei National Laboratory for Physical Sciences at the Microscale and Department of Modern Physics, University of Science and Technology of China, Hefei 230026, China}
\affiliation{
CAS Key Laboratory of Microscale Magnetic Resonance, University of Science and Technology of China, Hefei 230026, China}
\affiliation{
Synergetic Innovation Center of Quantum Information and Quantum Physics, University of Science and Technology of China, Hefei 230026, China}

\author{Xin Wang}
\affiliation{
Hefei National Laboratory for Physical Sciences at the Microscale and Department of Modern Physics, University of Science and Technology of China, Hefei 230026, China}
\affiliation{
CAS Key Laboratory of Microscale Magnetic Resonance, University of Science and Technology of China, Hefei 230026, China}
\affiliation{
Synergetic Innovation Center of Quantum Information and Quantum Physics, University of Science and Technology of China, Hefei 230026, China}

\author{Xinhua Peng}
\email[]{xhpeng@ustc.edu.cn}
\affiliation{
Hefei National Laboratory for Physical Sciences at the Microscale and Department of Modern Physics, University of Science and Technology of China, Hefei 230026, China}
\affiliation{
CAS Key Laboratory of Microscale Magnetic Resonance, University of Science and Technology of China, Hefei 230026, China}
\affiliation{
Synergetic Innovation Center of Quantum Information and Quantum Physics, University of Science and Technology of China, Hefei 230026, China}

\begin{abstract}
As a complementary analysis tool to conventional high-field NMR,
zero- to ultralow-field (ZULF) NMR detects nuclear magnetization signals in the sub-microtesla regime.
Spin-exchange relaxation-free (SERF) atomic magnetometers provide a new generation of sensitive detector for ZULF NMR.
Due to the features such as low-cost, high-resolution and potability,
ZULF NMR has recently attracted considerable attention in chemistry, biology, medicine, and tests of fundamental physics.
This review describes the basic principles, methodology and recent experimental and theoretical development of ZULF NMR,
as well as its applications in spectroscopy, quantum control, imaging, NMR-based quantum devices, and tests of fundamental physics.
The future prospects of ZULF NMR are also discussed.
\end{abstract}

\maketitle

\tableofcontents

\section{Introduction}\label{introduction}

Nuclear magnetic resonance (NMR), conventionally operated at high magnetic fields (on the order of tesla),
is a powerful analytical technique in chemistry, biology, and medicine~\cite{ernst1987principles, liang2000principles, wuthrich1986nmr, levitt2013spin}.
Although high magnetic fields for NMR have their virtues,
high-field $\textrm{NMR}$ spectrometers with superconducting magnets are expensive (on the order of $\$1$ million), require constant cryogenic maintenance, and are immobile.
Moreover, due to the existence of strong magnets,
high-field NMR can not be used on samples with metallic inclusions or implanted devices.
In magnetic resonance imaging (MRI)~\cite{liang2000principles},
the patients with pacemakers are not available to MRI, because the pacemakers can malfunction at strong magnetic fields.
Also, strong magnets often have severe magnetic field inhomogeneity,
limiting the ability to obtain high-resolution $\textrm{NMR}$ spectra.
Thus conventional NMR requires sophisticate magnetic field shimming systems to produce spatially uniform field.

In contrast to conventional high-field NMR,
an opposite trend called ``zero- to ultralow-field (ZULF) $\textrm{NMR}$"~\cite{weitekamp1983zero, blanchard2007zero, mcdermott2002liquid, ledbetter2011near,bevilacqua2009all, tayler2017invited, liu2013ultralow, garcon2019constraints}
has developed toward using ultralow, submicrotesla or even no external field at all.
\begin{figure*}[t]  
	\makeatletter
\centering
	\def\@captype{figure}
	\makeatother
	\includegraphics[scale=1.7]{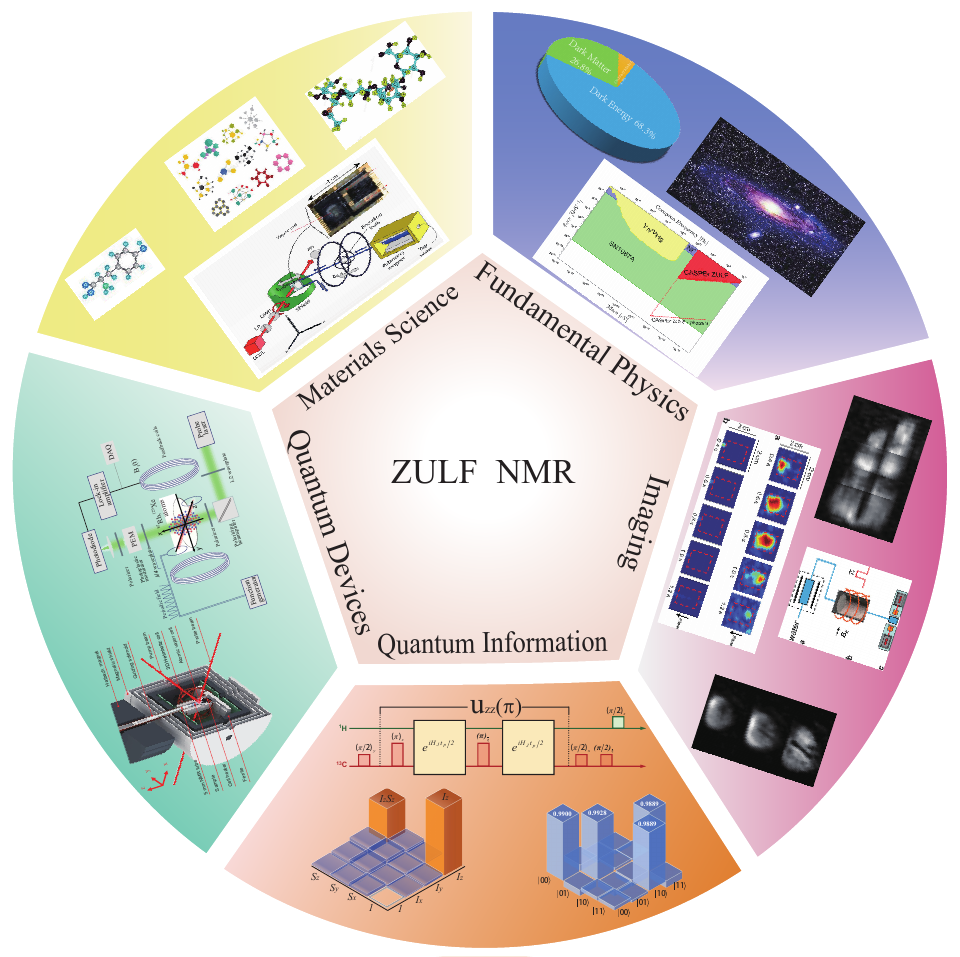}
	\caption{(Color online) Applications of ZULF NMR: (1) Materials science. Reprinted with permission from ref.~\cite{ledbetter2008zero}, Copyright @ 2008 National Academy of Sciences. (2) Imaging. Reprinted with permission from refs.~\cite{xu2006magnetic, savukov2013magnetic, savukov2013anatomical}, Copyright @ 2013 American Chemical Society,  @ 2006 National Academy of Sciences, @ 2013 Elsevier. (3) Quantum information processing. Reprinted with permission from ref.~\cite{jiang2018experimental}, Copyright @ 2018 American Association for the Advancement of Science. (4) NMR-based quantum devices. Reprinted with permission from ref.~\cite{jiang2019floquet}. (5) Tests of fundamental physics. Reprinted with permission from ref.~\cite{garcon2019constraints}, Copyright @ 2019 American Association for the Advancement of Science.}
	\label{figure1}
\end{figure*}
As shown in Fig.~\ref{figure1}, ZULF NMR has already demonstrated a variety of applications~\cite{ledbetter2008zero, xu2006magnetic, savukov2013magnetic, savukov2013anatomical, jiang2018experimental, jiang2019floquet}
ranging from materials science, quantum information processing, imaging, and NMR-based quantum devices to tests of fundamental physics.
Without the need of superconducting magnets,
ZULF NMR spectrometers could be inexpensive and portable,
and can be used to do research in situ without the need to deliver samples to the laboratory.
In addition to the economy and portability,
ZULF NMR has extremely high absolute magnetic field homogeneity and offers improved spectral resolution ranging from Hz all the way down to tens of mHz~\cite{blanchard2013high},
allowing one to detect certain low-frequency spin-spin interactions that would be challenging to observe in conventional high-field $\textrm{NMR}$.
For comparison,
one has to employ sophisticated and meticulously tuned shim coils to realize linewidths narrower than 0.5 Hz in high-field $\textrm{NMR}$~\cite{appelt2006chemical}.
Moreover,
ZULF NMR opens the door to new applications otherwise inaccessible to high-field $\textrm{NMR}$,
such as imaging and spectroscopy inside of metal objects or heterogeneous materials~\cite{xu2006magnetic, savukov2013magnetic, savukov2013anatomical}.
At ultralow field,
the Zeeman interactions are negligible or small enough that they can be treated as a perturbation on the internal spin-spin interactions (for example, J-coupling and dipole-dipole $\textrm{interactions}$).
This means that the nuclear spins are strongly coupled together compared with the coupling to the external magnetic field.
Such strong coupled spin systems should be particular interesting for quantum simulation,
such as investigating the topological order and Lee-Yang zeros of matter~\cite{luo2018experimentally, peng2015experimental}.


Although ZULF NMR has many advantages compared with high-field $\textrm{NMR}$,
its detection was challenging at the beginning of its development.
Because ZULF $\textrm{NMR}$ typically operates in low frequency,
the traditional Faraday induction coils are not highly sensitive for low-frequency signals;
thus the samples should be transferred to high field regime for detection at the early age of ZULF $\textrm{NMR}$~\cite{weitekamp1983zero}.
In recent years, with the advent of new quantum sensing techniques,
ZULF NMR gradually becomes mature and serves as a complementary tool to high-field $\textrm{NMR}$.
ZULF $\textrm{NMR}$ systems can use superconducting quantum interference devices (SQUIDs) as magnetic field sensors~\cite{mcdermott2002liquid, greenberg1998application, zotev2007squid}.
One drawback is that SQUIDs must operate at liquid-helium temperature and require constant cryogenic maintenance.
Recent years have seen increased developments in atomic magnetometers~\cite{budker2002resonant, budker2007optical, kominis2003subfemtotesla, li2006parametric, shah2007subpicotesla, fang2012situ, sheng2013subfemtotesla, RNGradiometer}, both in sensitivity and portability.
A spin-exchange relaxation-free (SERF) atomic magnetometer~\cite{allred2002high, dang2010ultrahigh}, with a measurement volume of $0.45~\textrm{cm}^3$,
has demonstrated sensitivity of $0.16~\textrm{fT}~{\rm{Hz}}^{-1/2}$, comparable to the most advanced SQUIDs.
SERF atomic magnetometers have been recently used in ZULF NMR to detect pure J-coupling spectra at zero-field~\cite{ledbetter2009optical, blanchard2013high,theis2011parahydrogen},
and determine spin-coupling topology at near-zero magnetic field~\cite{ledbetter2011near,jianginterference}.

One of the greatest challenges in ZULF NMR is how to selectively manipulate individual spin species.
This is the basis for developing more sophisticated pulse sequences, such as multidimensional NMR spectroscopy technique, dynamical decoupling, spin echo, etc.
Nuclear spins with different gyromagnetic ratios all have identical (zero) or near identical Larmor frequency at zero and ultralow field,
and thus individual manipulation of different spin species presents a challenge and limits the possible applications of ZULF NMR.
Previous examples of coherent isotropic averaging pulse sequences in zero-field $\textrm{NMR}$ have been limited to the homonuclear case~\cite{llor1995coherent,llor1995coherent2}.
More recent pulse sequences for heteronuclear spins have been demonstrated, but have been limited by the ability to generate individual arbitrary rotations for different spin species~\cite{thayer1986composite, lee1987theory, ledbetter2009optical, blanchard2013high, ledbetter2011near}.
We designed and performed a set of composite pulse sequences
that can realize universal quantum control of heteronuclear spin systems at zero and ultralow field~\cite{jiang2018experimental,bian2017universal}.
For manipulating homonuclear spin systems, a technique of shaped pulse sequences based on numerical optimization method was reported in ref.~\cite{jiang2018numerical}.
Tobias et al.~\cite{sjolander2020two, sjolander201713c} realized pulse sequences towards two-dimensional zero-field $\textrm{NMR}$,
which enables one to measure spin coupling networks.
To estimate the fidelity of pulse sequences,
several approaches~\cite{jiang2018experimental, ji2018time} have recently been demonstrated, including quantum information-inspired randomized benchmarking,
state tomography, and partial process tomography.
High-fidelity realization of such exquisite and universal pulse sequences in ZULF NMR is currently a major goal.

A wide range of applications based on ZULF $\textrm{NMR}$ have been demonstrated ranging from materials science, imaging, quantum information processing,
and NMR-based quantum devices to tests of fundamental physics.
For example, Barskiy et al. investigated the process of chemical exchange from ZULF NMR spectra~\cite{barskiy2019zero}.
Ledbetter et al.~\cite{RNLiquidstate} and Wu et al.~\cite{RNComagnetometer} demonstrated two prototypes of liquid-state comagnetometer based on ultralow-field NMR systems,
which show the potential to strengthen limits on spin-gravity coupling by several orders of magnitude.
More recently ZULF $\textrm{NMR}$ has been applied to searches for axion and axion-like-particle dark matter~\cite{garcon2019constraints,wu2019search}
and provides a promising axion research tool in the ultralight axion window.
In the theoretical researches, the absence of a large applied magnetic field allows for the measurement of antisymmetric spin-spin couplings~\cite{king2017antisymmetric},
which are related to chirality and have been proposed as a means for detecting molecular parity nonconservation.

In this review, we first focus on the basic principles and methodology used in ZULF NMR, such as main stages of $\textrm{NMR}$ experiments and atomic magnetometers.
Then, great potential applications of ZULF NMR are presented, such as high-resolution spectroscopy, quantum control, imaging, and the development of quantum devices.
Finally, we outline the future prospects of ZULF NMR in materials science, chemical analysis, and fundamental physics, etc.

\section{Basic principles}

We now describe the basic principles of ZULF NMR,
including the procedure of a ZULF NMR experiment and spin system at zero and ultralow field.

\subsection{Procedure of an NMR experiment}

\begin{figure*}[t]  
	\makeatletter
\centering
	\def\@captype{figure}
	\makeatother
	\includegraphics[scale=1.8]{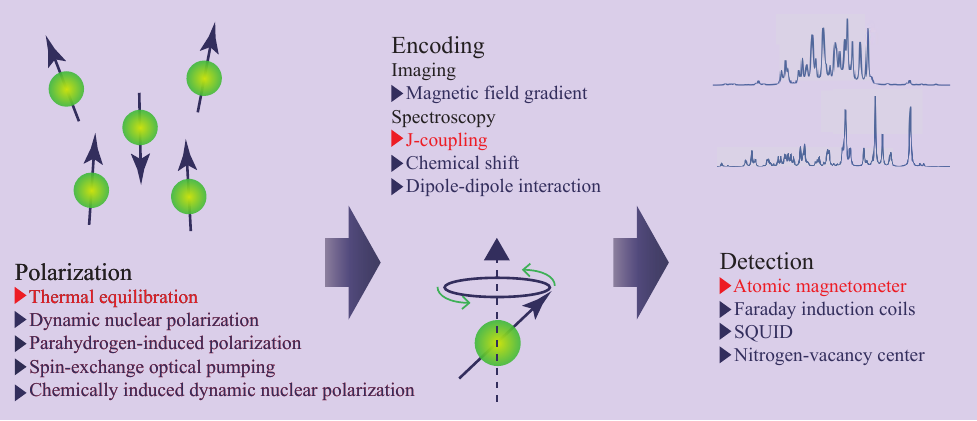}
	\caption{(Color online) Three stages of an NMR experiment - polarization, encoding, detection. In the presence of a strong magnetic field, the traditional method is that a sample's nuclear spins are polarized via thermal equilibration. Information about the structure or spatial distribution of nuclei is encoded in the form of spin resonances, such as chemical shifts or an applied magnetic field gradient; and Faraday induction coils are used to detect the NMR signal. At ultralow field, spin polarization can be achieved by dynamic nuclear polarization, spin-exchange optical pumping, chemically induced dynamic nuclear polarization or parahydrogen-induced polarization; encoding can take the form of J-coupling or dipole-dipole interaction; superconducting quantum interference devices (SQUIDs), atomic magnetometers or nitrogen-vacancy center can be used to detect spin resonances. Reprinted with permission from ref.~\cite{ledbetter2013nuclear}.}
	\label{procedure}
\end{figure*}

There are three main stages of an NMR experiment - polarization, encoding, and detection~\cite{liang2000principles, levitt2013spin, ledbetter2013nuclear},
as shown in Fig.~\ref{procedure}.
In conventional high-field \textrm{NMR}, the three main stages occur in a same spatial region,
whereas the three stages are usually spatially separated.
In traditional high-field \textrm{NMR}, a sample's nuclear spins are polarized thermally in a superconducting magnet.
In contrast,
ZULF NMR experiments usually use permanent magnet arrays (for example, low-cost Halbach magnets~\cite{tayler2017invited}) to polarize nuclear spins in samples.
Although the field strengths of permanent magnets (1-2 T) are relatively weaker than that of superconducting magnets,
they are low cost and portable.
In particular, they are not required with high field homogeneity.
However, the polarization obtained by thermal equilibration is on the order of 1 part per million (see Sec.~\ref{system}).
Alternatively, large nuclear spin polarizations can be achieved through hyperpolarization techniques,
such as parahydrogen-induced polarization (PHIP)~\cite{adams2009reversible, theis2011parahydrogen},
spin-exchange optical pumping (SEOP)~\cite{walker1997spin}, dynamic nuclear polarization (DNP)~\cite{maly2008dynamic},
and chemically induced dynamic nuclear polarization (CIDNP)~\cite{lawler1967chemically}.

Next, information about the samples' molecular structure is encoded in the form of spin-resonance frequencies.
In high-field NMR, the spin resonance frequencies depend on the applied magnetic fields according to spin gyromagnetic ratios (including the effect of chemical shift)
and internal spin-spin interactions.
In contrast, the encoding process is accomplished through internal spin-spin interactions such as J-coupling and dipole-dipole interactions.
The details are present in Sec.~\ref{system}.

The detection of NMR signals is traditionally accomplished with Faraday induction coils in high-field $\textrm{NMR}$,
in which the induced voltage is proportional to the rate of nuclear spin precession.
However, ZULF NMR typically operates in low frequency (below 1 kHz),
thus traditional Faraday induction coils are not highly sensitive.
With the advent of non-inductive sensors,
such as SQUIDs~\cite{mcdermott2002liquid, greenberg1998application, zotev2007squid} and atomic magnetometers~\cite{budker2007optical, kominis2003subfemtotesla, budker2002resonant, shah2007subpicotesla, RNGradiometer},
ZULF NMR gradually becomes a mature and useful technique.
In particular, atomic magnetometers based on spin-exchange relaxation-free mechanics have reached at the subfemtotesla level~\cite{kominis2003subfemtotesla, dang2010ultrahigh}.
In addition to sensitivity,
atomic magnetometers work best between room temperature and 180~$^0$C without the need of cryogenic condition,
and are well suited for detecting ZULF $\textrm{NMR}$ signals.
More recently,
it has realized a gradiometric ZULF $\textrm{NMR}$ spectrometer based on an atomic gradiometer~\cite{RNGradiometer},
which enables one order of magnitude enhancement in the signal-to-noise ratio (SNR) compared to the single-channel configuration.

\subsection{Spin systems at ultralow field}\label{system}

A liquid-state n-spin system at an external ultralow magnetic field can be described by the Hamiltonian:
\begin{equation}
 H = H_J+H_Z=\sum\limits_{i,j > i} {2\pi {J_{ij}}} {\textbf{I}_i} \cdot {\textbf{I}_j}- \sum\limits_j {{\gamma _j}{\textbf{I}_j} \cdot \textbf{B}},
\label{H}
\end{equation}
where $H_J$ and $H_Z$ are the scalar spin-spin coupling (so called J-coupling) and the Zeeman interaction respectively,
${J_{ij}}$ is the strength of the J-coupling between the $i$th and $j$th spins.
$\mathbf{I}_i = (I_{ix}, I_{iy}, I_{iz})$ is the spin angular momentum operator of the $i$th spin,
and the reduced Planck constant $\hbar$ is set to one.
The effects of dipole-dipole interactions are ignored here, as in an isotropic liquid,
they are averaged out due to the rapid and random flipping motion of the molecules.
In ZULF $\textrm{NMR}$,
the Zeeman interactions are much smaller than the coupling interaction ($|\gamma _j\textbf{B}| \ll|J|$),
so that Zeeman interactions are treated as perturbations.
As we show below, the spectra of samples at a near-zero field usually split into, for example, doublet, triplet, etc~\cite{ledbetter2011near, appelt2010paths, appelt2007phenomena}.
Such near-zero-field splitting patterns provide plentiful information on energy degeneracies and can be used as a diagnostic tool to determine molecular structure.

\begin{figure*}[t]  
	\makeatletter
\centering
	\def\@captype{figure}
	\makeatother
	\includegraphics[scale=0.53]{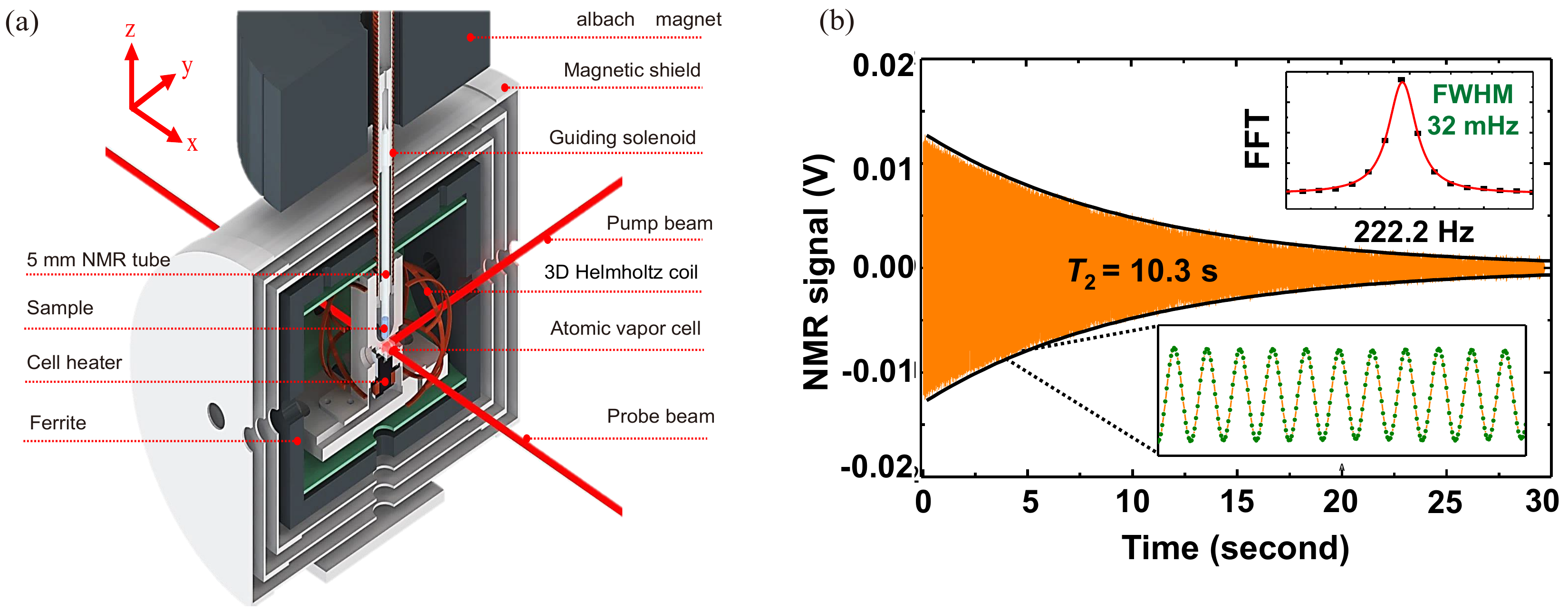}
	\caption{(Color online) (a) Experimental apparatus of ZULF NMR based on an atomic magnetometer. The NMR sample is contained in a standard 5-mm NMR tube and pneumatic shuttled between a Halbach magnet and detection region. (b) Single-shot zero-field NMR signal of $^{13}$C-formic acid. The lifetime of the singlet-triplet coherence is 10.3s, and the FWHM (full width at half maximum) is 32 mHz.  Reprinted with permission from ref.~\cite{jiang2018experimental}, Copyright @ 2018 American Association for the Advancement of Science.}
	\label{setup}
\end{figure*}

We consider the thermal equilibration in a permanent magnet,
which is a universal approach for preparing initial spin polarization of samples.
The schematic diagram of ultralow-field NMR apparatus (see, e.g., refs.~\cite{jiang2018experimental,ji2018time,RNGradiometer}) based on an atomic magnetometer is shown in Fig.~\ref{setup}a.
Liquid-state samples are contained in standard 5-mm $\textrm{NMR}$ tubes,
and pneumatically shuttled between a prepolarizing Halbach magnet (the field in the center is about 1-2~T) and an alkali-metal vapor cell.
During the transfer, a guiding magnetic field is applied along the transfer direction (along the $z$ axis).
The magnitude of the guiding field ($B_g\approx 10^{-4}~\textrm{T}$) satisfies $B_g \gg |J_{ij} /(\gamma_i-\gamma_j)|$,
so the spin state remains the high-field equilibrium state.

The way that the guiding field is switched off plays a crucial role in determining the initial state and,
in turn, the amplitude and phase of the oscillating NMR magnetization signal produced~\cite{jiang2018experimental,blanchard2007zero}.
We consider two cases where the guiding field is switched off suddenly or adiabatically.
When the guiding field is suddenly switched off,
the relevant initial spin state is
\begin{equation}
\begin{array}{l}
\rho_0=\mathbbm{1}/2^n-\sum_j \varepsilon_j {I}_{jz},
\end{array}
\label{rho0}
\end{equation}
where $\varepsilon_j=\gamma_jB_p/k_\textrm{B} T \sim 10^{-6}$,
${I}_{jz}$ is the spin angular momentum operator along the $z$ orientation of the $j$th spin with gyromagnetic ratio $\gamma_j$,
$k_\textrm{B}$ is the Boltzmann constant,
$T$ is the temperature.
In this case, $\rho_0$ is called as sudden state.
Alternatively, when the guiding field is adiabatically turned off (with the characteristic time scale defined by the strength of the J-coupling),
the populations at high field are converted to the populations of the zero- or ultralow-field eigenstates.
For example, the initial state of a two-spin system ($I$ and $S$) is~\cite{emondts2014long, jiang2018experimental, blanchard2007zero}
\begin{equation}
\begin{array}{l}
\rho_0=\frac{(\gamma_I+\gamma_S)B_p}{8k_BT}(I_z+S_z)- \frac{(\gamma_I-\gamma_S)B_p}{4k_BT}(I_xS_x+I_yS_y),
\end{array}
\label{rho02}
\end{equation}
which is the adiabatic state.
One can similarly achieve the form of the adiabatic state of multi-spin molecules.

The nuclear spins in the sample then evolve at zero or ultralow bias magnetic field and generate an oscillating magnetization,
\begin{equation}
\rho(t)=e^{-iHt}\rho_0 e^{iHt},
\end{equation}
where $\rho(t)$ is the time dependent density matrix.
The magnetization along $\eta=x,y,z$ can be calculated by
\begin{equation}
M_{\eta}(t)=n\textrm{Tr} [\rho(t) \sum_j \gamma_j  I_{j,\eta}],
\label{M}
\end{equation}
where $n$ is the molecular density.
Expanding Eq.~\ref{M} in terms of matrix elements between the normalized eigenstates $\left\{\ket{\phi_k } \right\} $ of $H$:
\begin{equation}
	M_{\eta}(t)=n\sum_{a,b,j} \rho_{ab} \bra{\phi_b}\left(\gamma_j I_{j\eta} \right) \ket{\phi_a} e^{ -i2 \pi \nu_{ab} t},
\end{equation}
where $\ket{\phi_a}$ and $\ket{\phi_b}$ denote different eigenstates, $\nu_{ab}$ is the transition frequency between the eigenstates $\ket{\phi_a}$ and $\ket{\phi_b}$ and $\rho_{ab}=\bra{\phi_a}\rho_0\ket{\phi_b} $.
Further, the nuclear magnetization can produce magnetic fields on the vapor cell of atomic magnetometer.
Approximating the $\textrm{NMR}$ sample as a sphere,
the magnetic moment due to the sample magnetization is $\mathbf{m}=(4\pi r_0^3/3)\mathbf{M}$,
where $r_0$ is the radius of the sample, and $\mathbf{M}=[M_x, M_y,M_z]$ is the nuclear magnetization.
The magnetic field originated from the sample at the location of the atomic magnetometer is~\cite{RNGradiometer}
\begin{align}
\mathbf{B}_s=\frac{\mu_{0}}{4\pi}\frac{3\hat{\mathbf{n}}(\mathbf{m}\cdot \hat{\mathbf{n}})-\mathbf{m}}{r^3},
\end{align}
where $\mu_0$ is the permeability of vacuum, $\hat{\mathbf{n}}$ is the unit vector pointing from the sample to the magnetometer and $r$ is the distance between them.
After applying a fast Fourier transform to the time-domain magnetometer signal,
frequency-domain ZULF NMR spectra can be achieved.
It is worthy noting that the magnetometer is sensitive to the oscillating magnetic field along the $z$-axis (see Sec.~\ref{subsecexp}), so only $z$-magnetic field component is considered.

\section{Experimental apparatus}\label{subsecexp}

The schematic diagram of zero- to ultralow-field NMR apparatus based on an atomic magnetometer is shown in Fig.~\ref{setup}.
Atomic magnetometers usually operate at zero or ultralow magnetic field and provide an exquisite sensitive approach for detecting ZULF NMR signals.
The atomic magnetometer consists of a warm alkali-metal vapor $\textrm{cell}$ (for example, $^{87}$Rb vapor $\textrm{cell}$).
Nitrogen (N$_2$) with a pressure of about one atmosphere is added as buffer gas in the vapor cell,
and improve the detection bandwidth (above 100 Hz) of atomic magnetometer.
It is necessary to match the dimensions of the vapor cell and the detected sample.
The typical size of vapor cells is 0.5-1 cm, comparable to the standard 5-mm NMR tube.
We should note that the conventional atomic magnetometers use large vapor cells,
which are usually several centimeter, and thus are not applicable to ZULF NMR detection.
The vapor cell works in a stable temperature 180~$^{\textrm{o}}$C by resistive heating with using twisted coils and high-frequency AC current,
and is placed inside a five-layer mu-metal magnetic shield (shielding factor of $10^6$).
The alkali-metal atoms in the vapor cell are optically pumped with a circularly polarized laser beam propagating in the $y$ direction.
The pump-laser frequency is tuned close to the center of the buffer-gas (N$_2$) broadened and shifted D1 line.
The magnetic field is measured via optical rotation of linearly polarized probe laser beam propagating along the $x$ direction (for example, see ref.~\cite{allred2002high}).
The frequency of the probe laser is detuned from the D2 transition by about 100~GHz.
To suppress the influence of low-frequency noise,
the polarization of the probe laser beam is usually modulated by a photoelastic modulator (PEM)~\cite{RNGradiometer} or Faraday modulator~\cite{fang2012situ},
and the signal is demodulated by a lock-in amplifier.

We now calculate the oscillatory signal of an atomic magnetometer.
The precise description of the interaction between light and atoms requires the use of a density matrix, but the ground state of alkali-metal atoms can be described by a Bloch equation for the electron spin polarization $\textbf{P}$~\cite{allred2002high, RNGradiometer, jianginterference}:
\begin{equation}
	\frac{d \textbf{P}}{dt}= \frac{1}{q}\left[\gamma_{e} \textbf{B}(t) \times \textbf{P} +R_{op} \left(\hat{ \textbf{y}}-\textbf{P} \right)  -R_{rel}\textbf{P} \right],
\label{bloch}
\end{equation}
where $\textbf{P} = \left\langle \textbf{S} \right\rangle /S$, $\textbf{S}$ is the electron spin angular momentum, $\gamma_e=g_s\mu_B$, $g_s \approx 2$ is the electron Land\'{e} factor, $\mu_B$ is the Bohr magneton, $\textbf{B}(t)$ is the applied magnetic field, $R_{op}$ is the optical pumping rate due to the pump beam, $\hat{ \textbf{y}}$ is the direction of the pump beam, $R_{rel}$ is the spin relaxation rate in the absence of optical pumping, and $q$ is the nuclear slowing-down factor.
In the present of a slowly changing magnetic field, the quasi-steady-state solution to Eq.~\ref{bloch} can be found,
we introduce $\beta = \gamma_{e} \textbf{B}/ \left(R_{op} + R_{rel}\right)$ and $P_0=R_{op}/\left(R_{op} + R_{rel}\right)$. The solution of above equation is 
\begin{eqnarray}
	\begin{cases}
		P_x&=P_0\frac{\beta_z +\beta_x\beta_y }{1+|\beta|^2},\\
		P_y&=P_0 \frac{1+\beta_y^2}{1+|\beta|^2},\\
		P_z&=P_0 \frac{-\beta_x+ \beta_z\beta_y}{1+|\beta|^2}.\\
	\end{cases}	
\end{eqnarray}
Because the propagation direction of the probe light is along $x$, the Faraday rotation angle of the probe light is proportional to $P_x$~\cite{allred2002high, dang2010ultrahigh}.
When the magnetic field is very small ($|\beta|\ll 1$),
\begin{equation}
	P_x  \approx P_0 \beta_z.
\end{equation}
Hence, atomic magnetometer is primarily sensitive to magnetic field along $z$ direction. However, in the presence of a bias magnetic field along $y$, 
\begin{equation}
	P_x  \approx P_0 \frac{\beta_z+\beta_x \beta_y}{1+\beta_y^2}.
\end{equation}
In this situation, atomic magnetometer can be simultaneously sensitive to the $x$ and $z$-direction magnetic fields.
As discussed in Sec.~\ref{asym}, an asymmetric ZULF NMR spectroscopy can be generated due to that atomic magnetometer is both sensitive to the NMR signals along $x$ and $z$. A detailed discussion can be found in ref.~\cite{jianginterference}.

Samples ($\sim 200$~$\mu$L) were flame-sealed under vacuum in a standard $5$-mm glass NMR tube following five freeze-pump-thaw cycles in order to remove dissolved oxygen,
which is otherwise a significant source of relaxation.
Liquid-state samples are pneumatically shuttled between a prepolarizing Halbach magnet and an alkali-metal vapor cell.
The bottom of the $\textrm{NMR}$ sample tube is $\sim$1 mm above a vapor cell of the atomic magnetometer.
Before detection, it is usually necessary to apply magnetic field pulses on the samples' spins (see Sec.~\ref{control}) for optimizing NMR signal amplitude.
There are two sets of orthogonal three dimensional coils placed around the vapor cell.
One set of coils is used to apply uniform bias magnetic field.
Another set of coils is used to apply pulses.
To apply pulses,
the desired waveform is written as a list of time/voltage coordinates ($t$,$V_x$,$V_y$,$V_z$),
which are converted into a three-channel analog output voltage.
Each output channel is fed into a linear high-power amplifier,
whose output is connected to one of the coils.

\begin{figure*}[t]  
	\makeatletter
\centering
	\def\@captype{figure}
	\makeatother
	\includegraphics[scale=0.55]{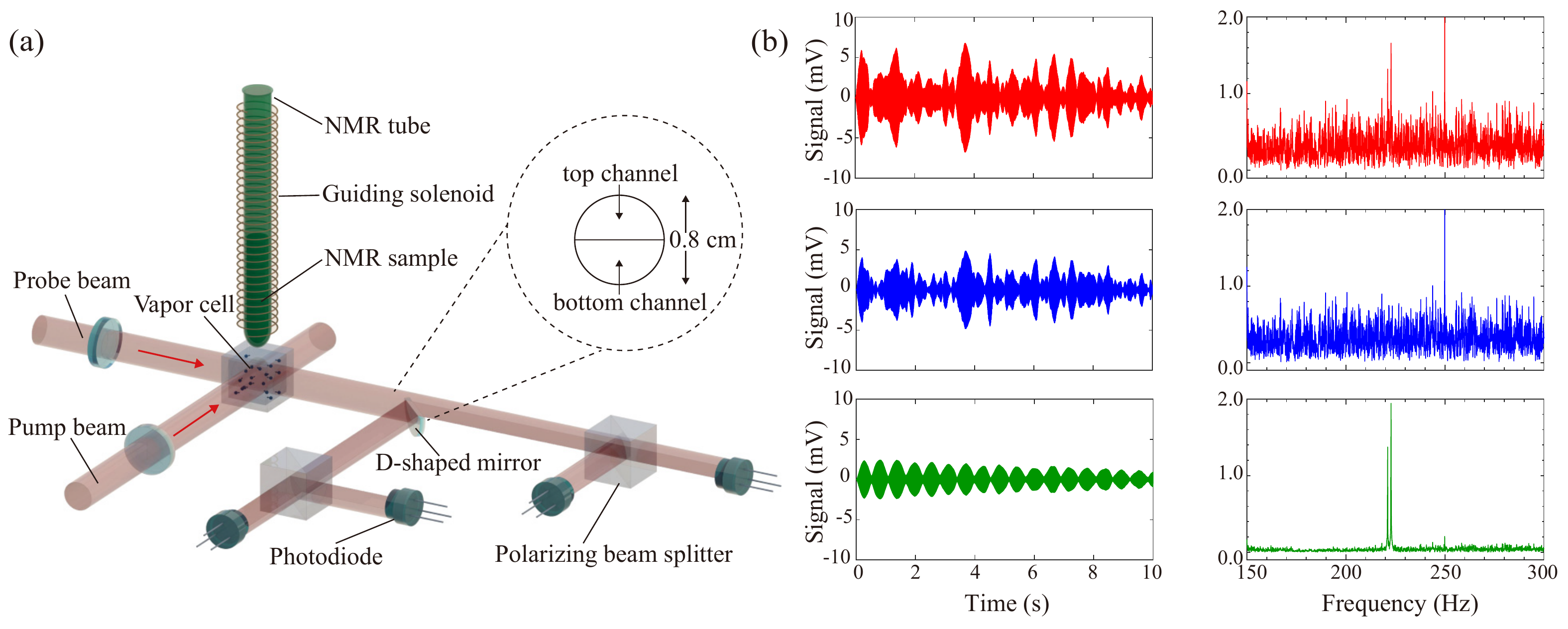}
	\caption{(Color online) (a) Schematics of the gradiometric detection for ZULF NMR. The detection is based on an atomic magnetic gradiometer. Inset shows that the probe beam (with diameter $\sim 0.8~\textrm{cm}$) is split into top and bottom channels using a D-shaped mirror. (b) Single-shot ultralow-field NMR of $^{13}$C-formic acid. The $^{13}$C-formic acid time-domain signals and their spectra are observed in top (top), bottom (middle), and gradiometric (bottom) channels in the presence of a small magnetic field. Reprinted with permission from ref.~\cite{RNGradiometer}, Copyright @ 2019 American Physical Society.}
	\label{setup-2}
\end{figure*}

Current ZULF NMR spectrometers are always equipped with high-quality magnetic shields to ensure that ambient magnetic field noise does not dwarf the magnetization signal.
An alternative approach is to separate the magnetization signal from the noise based on their differing spatial profiles, as can be achieved using a magnetic gradiometer.
We have recently realized a gradiometric ZULF NMR spectrometer~\cite{RNGradiometer} (see Fig.~\ref{setup-2})
with a magnetic gradient noise of 17$~\textrm{fT}~\textrm{cm}^{-1} ~ \textrm{Hz}^{-1/2}$ in the frequency range of 100-400 Hz,
based on a single vapor cell (0.7$\times$0.7$\times$1.0 $\textrm{cm}^3$).
The gradiometric spectrometer achieves 13-fold enhancement in the signal-to-noise ratio (SNR) compared to the single-channel configuration.
By reducing the influence of common-mode magnetic noise, the gradiometric $\textrm{NMR}$ spectrometer enables the use of compact and low-cost magnetic shields.

\section{Spectroscopy}

Due to the high absolute magnetic field homogeneity,
ZULF NMR can realize ultrahigh-resolution spectroscopy as a powerful diagnostic tool for non-destructive structural investigations of matter.
This is contrast to high-field NMR, where magnetic field inhomogeneity is the main factor to cause NMR line broadening.
In addition, the spectral linewidth of ZULF NMR depends on spin relaxation times.
In contrast to high-field case, the absence of certain relaxation pathways~\cite{emondts2014long} (for example, chemical shift anisotropy) allows ZULF NMR to achieve long spin relaxation times. 
For example, nuclear spin relaxation time is about $10.3$~s for formic acid (e.g., see Fig.~\ref{setup}b)~\cite{jiang2018experimental}, $4.7$~s for acetonitrile, $8.8$~s for acetic acid, and $6.9$~s for fully labeled acetonitrile.
Some NMR samples may have short relaxation times. For example, nuclear spin relaxation time is about $0.8$~s for formaldehyde.
Overall, ZULF NMR offers spectral resolution ranging from Hz all the way down to tens of mHz.
Moreover, the absence of truncation by a large applied magnetic field means that ZULF NMR is capable of measuring spin-dependent interactions that do not commute with the Zeeman Hamiltonian~\cite{blanchard2015measurement},
which is not generally accessible in conventional high-field NMR experiments.
In this section,
we review some recent experiments of ZULF NMR spectroscopy.

\begin{figure*}[t]  
	\makeatletter
\centering
	\def\@captype{figure}
	\makeatother
	\includegraphics[scale=1]{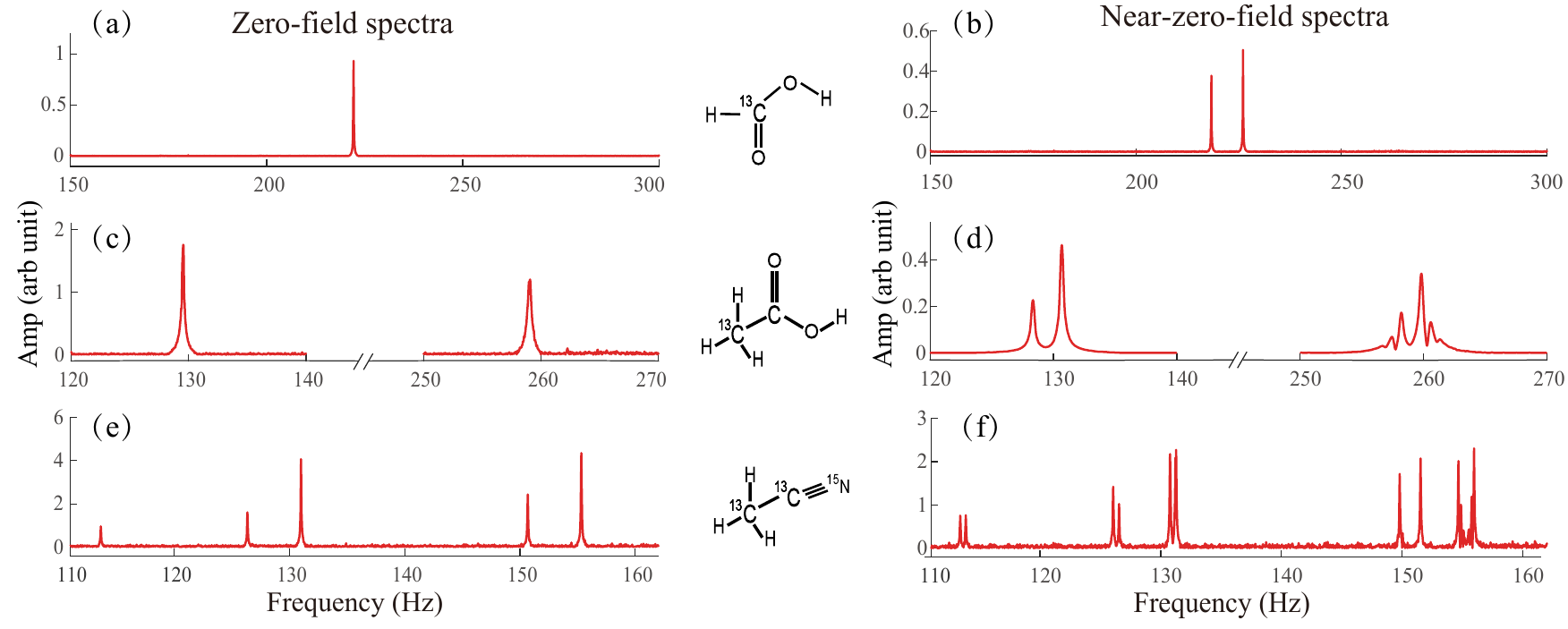}
	\caption{(Color online) Experimental ZULF NMR spectra. The initial spin state is prepared by the guiding field along the $z$ axis. (a, b) Formic acid (H$^{13}$COOH, where the acidic proton is negligible due to rapid exchange), (c, d) Formaldehyde (H$_3$$^{13}$CCOOH), (e, f) Acetonitrile ($^{13}$CH$_3$$^{13}$C$^{15}$N). The spectra are detected in the absence of a magnetic field and the presence of a magnetic field.}
	\label{ZQ&SQ}
\end{figure*}

\subsection{Zero-field NMR spectroscopy}\label{ZFNMR}

Zero-field NMR spectra are entirely determined by J-couplings between nuclear spins,
which are thus called ``pure J-coupling spectroscopy''.
J-couplings, which are mediated through chemical bonds connecting two spins,
can impart valuable information about chemical bonding and molecular structure.
Typically, the strength of J-couplings is in the range of 100-200~Hz for proton and carbon nuclei separated by a single covalent bond,
and it quickly decreases as the number of intermediate bonds grows~\cite{ernst1987principles}.
The high-resolution of zero-field $\textrm{NMR}$ spectra could be at the level of 10~mHz~\cite{ledbetter2009optical, blanchard2013high, ledbetter2013nuclear} and facilities the measurement of multi-bond J-coupling.
Here we summarize the recent development of zero-field NMR spectroscopy and its applications in chemical fingerprinting~\cite{ledbetter2009optical}.
We consider typical XA$_n$ type molecules,
where each ``X'' spin couples to $n$ equivalent ``A'' spins with the same J-coupling strength J$_{\textrm{AX}}$.
The J-coupling Hamiltonian is written as 
\begin{equation}
	H_J=2\pi J_{AX}\mathbf{K_A} \cdot \mathbf{S},
\end{equation}
here $\mathbf{K_A}$ is sum of the spin $A$, $\mathbf{S}$ is the angular momentum of spin $X$, and the total angular momentum $\mathbf{F_A} = \mathbf{K_A} + \mathbf{S}$.
The eigenstates $\ket{{f_A,{m_{f_A}}}}$ have energy
\begin{equation}
	E^{(0)} =J_{AX}/2 [f_A(f_A + 1) - k_A(k_A + 1) -s(s +1)],
\end{equation}
and its degeneracy is $2f_A+1$.
Because the detectable magnetization is $M_z(t)=n \textrm{Tr}[\rho(t) \sum \limits_j \gamma_j I_{jz}]$,
we obtain selection rules for observable coherences: $\Delta f_A =0,\pm 1$ and $\Delta m_{f_A} = 0$.
The equivalence of the XA$_n$ imposes an additional selection rule, $\Delta k_A = 0$~\cite{ledbetter2011near, appelt2010paths}.
Accordingly, one could observe $n-2k$ resonance lines with the frequencies $\nu^0=\frac{1}{2}J_{AX}(1+n-2k)$, where $k=n/2-k_A$.
Under these circumstances, we can conclude that for odd $n$,
an XA$_n$ system produces $(n+1)/2$ lines at frequencies ranging from $J_{AX}$ to $(n+1)/2 \times J_{AX}$ in step of $J_{AX}$,
and for even $n$ the number of lines is $n/2$ with frequencies ranging from $3/2J_{AX}$ to $(n+1)/2 \times J_{AX}$ in step of $J_{AX}$.
As shown in Fig.~\ref{ZQ&SQ}a, c,
the resulting zero-field J-coupling spectra consist of a single line at J for XA, and two lines,
one at J and the other at 2J, for XA$_3$.

A variety of chemical samples have been detected at zero field,
and their zero-field spectra are high-resolution and information-rich spectra that are well suited for chemical fingerprinting.
For XA$_n$ system, such as formic acid (H$^{13}$COOH),
formaldehyde ($^{13}$CH$_2$O),
methanol ($^{13}$CH$_3$OH),
acetonitrile ($^{13}$CH$_3$CN),
acetone((CH$_3$)$^{13}$CO{CH$_3$}),
trimethyl phosphate (PO(OCH$_3$)$_3$))~\cite{blanchard2007zero,ledbetter2011near,mcdermott2002liquid}.
For (XA$_n$)B$_m$ system, such as methyl formate(H$^{13}$COOCH$_3$), formamide (H$_2^{15}$NOH), ethanol-1 (CH$_3^{13}$CH$_2$OH), ethanol-2 ($^{13}$CH$_3$CH$_2$OH),glycerol-2 (HO$^{13}$CH(CH$_2$OH)$_2$)~\cite{ledbetter2009optical,blanchard2007zero}.
For more complex systems: fully labeled acetonitrile ($^{13}$CH$_3$$^{13}$C$^{15}$N) (see Fig.~\ref{ZQ&SQ}e)~\cite{ledbetter2011near, RNGradiometer},
acetone doubly labeled (H$_3$$^{13}$CCO$^{13}$CH$_3$)~\cite{weitekamp1983zero}, aromatic compounds~\cite{blanchard2013high}.

\subsection{Near-zero-field NMR spectroscopy}

Although zero-field NMR spectra can distinguish some chemical groups,
it still leaves some ambiguity in determination of chemical groups,
and this ambiguity can be removed by application of small magnetic fields~\cite{ledbetter2011near, appelt2010paths, appelt2007phenomena}.
Here we introduce the investigations of near-zero-field $\textrm{NMR}$ spectroscopy,
where the Zeeman interaction can be treated as a perturbation to J-couplings.
For simplicity, the direction of the applied bias field is chosen as the quantization axis ($z$) and the sensitive axis of the atomic magnetoemter is changed to $y$.
The presence of very small magnetic fields results in splitting of the zero-field NMR lines (Fig.~\ref{ZQ&SQ}b, d, f),
imparting considerable additional information to the pure zero-field spectra.
The molecular conformations of XA and XA$_2$ cannot be distinguished because the zero-field NMR spectra of XA and XA$_2$ both show single NMR peak (see Sec.~\ref{ZFNMR}).
In contrast, near-zero-field NMR spectra of XA and XA$_2$ are significantly different.
We take XA$_n$ spin system as an example to analyse.
The system Hamiltonian is described in Eq.~\ref{H}.
To first order in $B_z$, eigenstates are those of the unperturbed Hamiltonian,
and Zeeman shifts of the eigenvalues can be read from the diagonal matrix elements of the Zeeman perturbation.
One finds~\cite{blanchard2007zero, ledbetter2011near, appelt2010paths, appelt2007phenomena}
\begin{equation}
	\begin{split}
	&\Delta E(f_A, k_A, m_{f_A})\\
	&=-\bra{{f_{A}m_{f_A}}} B_z(\gamma_A K_{Az} + \gamma_X S_z) \ket{ {f_{A}m_{f_A}}}\\
	&=-B_z\sum\limits_{m_{kA},m_s} \left\langle {k_{A}sm_{k_A}m_s|f_{A}m_{f_A}} \right\rangle^2(\gamma_A m_{k_A} + \gamma_X m_s),
\end{split}
\end{equation}
where $\gamma_A$ and $\gamma_X$ are the spins A and X gyromagnetic ratios respectively, and $\left\langle k_{A}sm_{kA}m_s|{f_{A}m_{f_A}} \right\rangle$ is the Clebsch-Gordan coefficient.
The signal in our experiment is the $y$ component of the magnetization
$M_y(t)=n \textrm{Tr}[\rho(t) \sum \limits_j \gamma_j I_{jy}]$,
the relevant selection rules are $\Delta f_A =0,\pm 1$, $\Delta k_A = 0$, and $\Delta m_{f_A}= \pm 1$.
Accordingly, one could observe $2(n-2k)$ resonance lines centered at $\nu^0$, with the frequencies
\begin{equation}
	\begin{cases}
\nu_{f_A,m_{f_A};k_A}^{f_A',m_{f_A}\pm1;k_A}=\nu^0 +\Delta \nu,\\
\nu^0 =\frac{1}{2}J_{AX}(1+n-2k),\\
\Delta \nu=[\frac{2m_{f_A} (-\gamma_A+\gamma_X)}{1+n-2k} \pm \frac{(n-2k)\gamma_A+\gamma_X}{1+n-2k}]B_z,
	\end{cases}
\end{equation}
where $\nu^0$ is frequency of resonance line in zero field (see Sec.~\ref{ZFNMR}) and $\Delta \nu$ implies  Zeeman splitting.
For example of $n=1$, a doublet (see Fig.~\ref{ZQ&SQ}b) for the transition between states with $f_A=1$ and $f_A=0$:
\begin{equation}
	\nu_{0,0}^{1,\pm1}= J\pm B_z(\gamma_A+\gamma_X)/2.
\end{equation}
More examples can be found in refs.~\cite{ledbetter2011near, appelt2010paths}.
Some samples have experimentally detected near-zero-field NMR spectra,
such as formic acid (H$^{13}$COOH), formaldehyde ($^{13}$CH$_2$O), methanol ($^{13}$CH$_3$OH), acetonitrile ($^{13}$CH$_3$CN), fully labeled acetonitrile ($^{13}$C{H$_3$}$^{13}$C$^{15}$N) \cite{ledbetter2011near,mcdermott2002liquid,RNGradiometer}.

\begin{figure}[t]  
	\makeatletter
\centering
	\def\@captype{figure}
	\makeatother
	\includegraphics[scale=0.5]{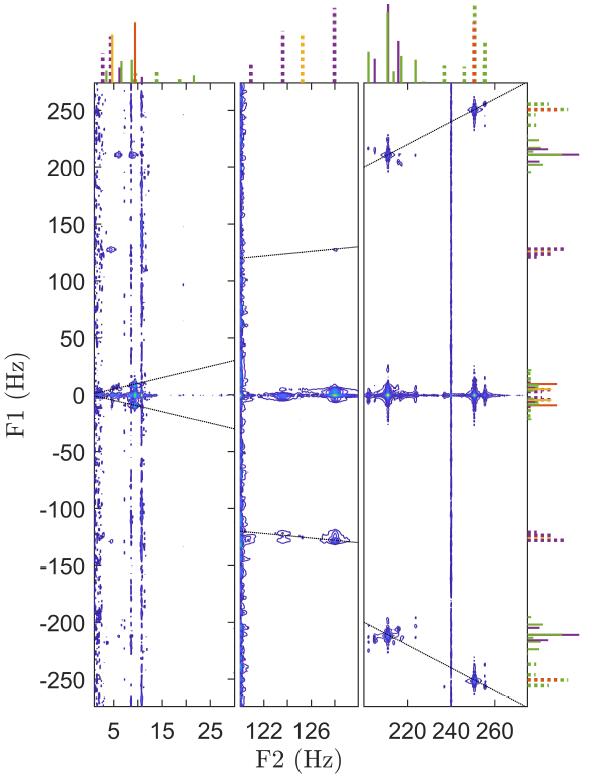}
	\caption{(Color online) Experimental two-dimensional zero-field NMR spectrum of ethanol mixture using ($\pi_{x/y}-t_1-\pi_{x/y}$) pulse sequence. The entire spectral window (both positive and negative frequencies) is displayed in F1, whereas only the positive part of the F2 axis is shown. The spectrum contains cross-peaks between transitions belonging to the same spin-state manifolds. Meanwhile, there is no coherence transfer, and therefore no cross-peaks, between either the two isotopomers or peaks corresponding to the same isotopomer but different spin-manifolds. Reprinted with permission from ref.~\cite{sjolander2020two}.}
	\label{2D}
\end{figure}

\subsection{Two-dimensional NMR spectroscopy}

Spectral complexity increases rapidly with spin-system size.
The development of two-dimensional spectroscopy is a major reason behind the analytical power of NMR, allowing the resolution of more crowded spectra.
Additionally, many pulse sequences exist that enable the mapping of coupling networks, the simplification of spectral assignment, and structure elucidation.
In liquid-state analytical chemistry, multiple-quantum coherence filters combined with two-dimensional detection techniques provide one of the standard ways to map coupling networks~\cite{ernst1987principles}.
Recently, Sjolander et al.~\cite{sjolander2020two, sjolander201713c} introduced a technique of two-dimensional correlation and single- and multiple-quantum experiments in the context of liquid-state zero-field J-spectroscopy.
For example, at zero field the spectrum of ethanol appears as a mixture of $^{13}$C isotopomers, and correlation spectroscopy is useful in separating the two composite spectra,
as shown in Fig.~\ref{2D}.
Two-dimensional spectroscopy further improves the high resolution attained in zero-field NMR since selection rules on the coherence-transfer pathways allow for the separation of otherwise overlapping resonances into distinct cross-peaks~\cite{sjolander2020two}.

\subsection{Asymmetric NMR spectroscopy}
\label{asym}

\begin{figure}[t]
    \centering
    \includegraphics[width=\linewidth]{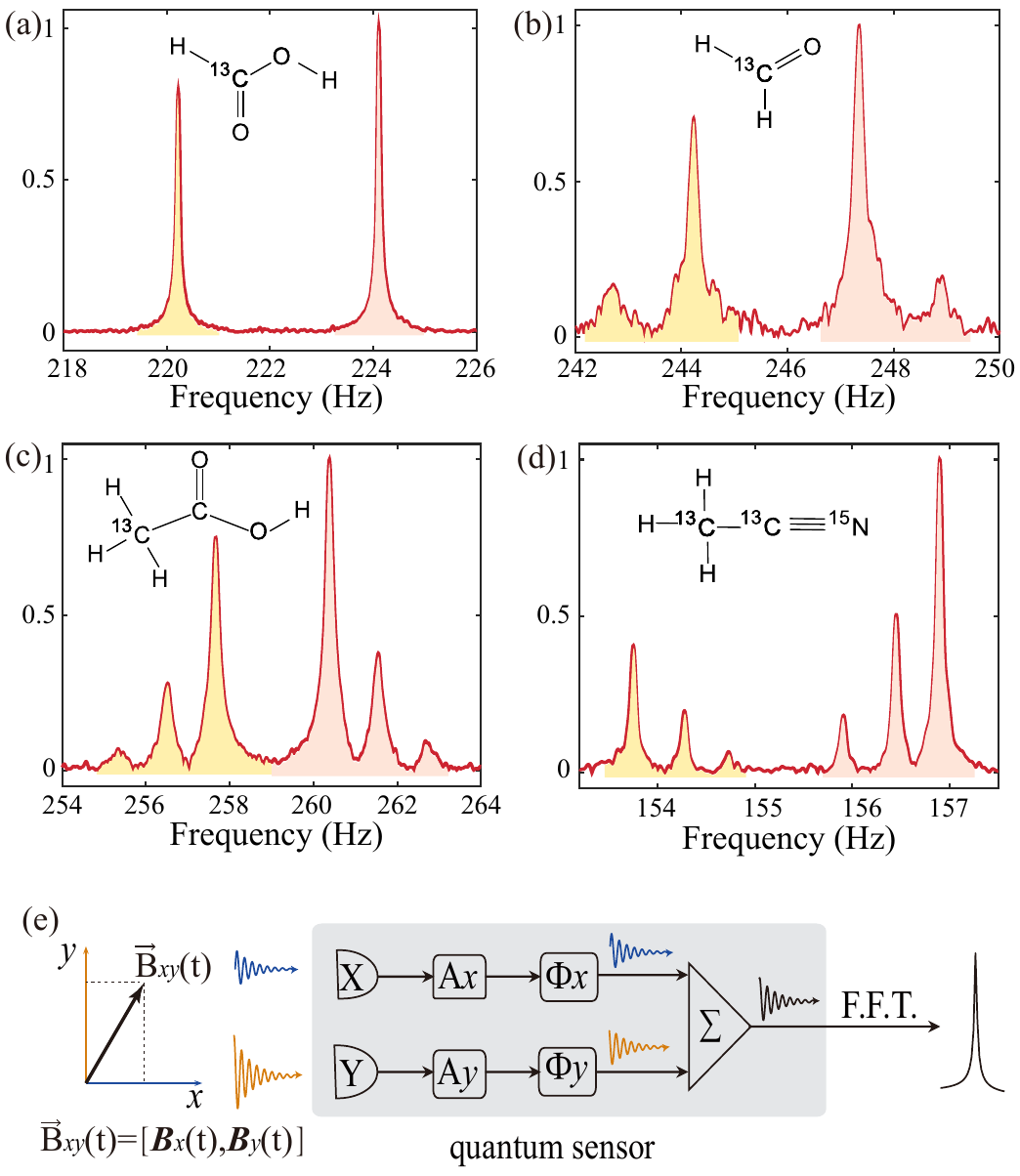}
    \caption{(Color online) Experimental asymmetric ultralow-field NMR spectra. (a) Formic acid, (b) Formaldehyde, (c) Acetic acid, (d) Fully labeled acetonitrile. These are partial spectra of the samples. A small magnetic field ($\approx$$72.9$~nT) is applied along $x$ in the experiments. Results show that the amplitudes of the $\textrm{NMR}$ peaks with yellow color are different from the corresponding peaks with red color. (e) Basics of the quantum sensor resulting in the interference effect: The magnetic field components along the $x$- and $y$-axes both affect the sensor but the observed signal is given only by the sum of their contributions, which have different sensitivities (amplitudes $\textrm{A}_x$, $\textrm{A}_y$) and phases ($\Phi_x$, $\Phi_y$). Reprinted with permission from ref.~\cite{jianginterference}, Copyright @ 2020 Wiley-VCH GmbH.}
    \label{Fig-Asymmetry}
\end{figure}

It has been extensively demonstrated that the ultralow-field NMR signals recorded with atomic magnetometers
uncommonly differ from the expected values with a few tens of percent distortion~\cite{ledbetter2011near, RNComagnetometer, RNLiquidstate, RNGradiometer, garcon2019constraints, wu2019search}.
Ultralow-field NMR spectra of even simple samples like formic acid (containing $^{13}$C-$^1$H spin pairs) suffer from severe asymmetric amplitudes (Fig. \ref{Fig-Asymmetry}),
differing greatly from those predicted by the standard $\textrm{NMR}$ theory~\cite{appelt2010paths, appelt2007phenomena, ledbetter2011near}.
Figure~\ref{Fig-Asymmetry} shows the experimentally observed NMR spectra of some samples, clearly exhibiting asymmetric characteristics on their spectral amplitude.
For example, the relative NMR peak amplitudes for acetic acid are theoretically predicted to be 1:3:6:6:3:1,
but in practical experiments the ratio is measured to be 1:4.6:12.3:16.4:6.2:1.5 (as shown in Fig.~\ref{Fig-Asymmetry}c).

Recently, we uncovered an interference effect in atomic magnetometers~\cite{jianginterference},
which causes an important systematic effect to greatly deteriorate the accuracy of measuring magnetic fields.
Based on the reported interference effect, good agreement is found between the prediction and the asymmetric amplitudes of resonant lines in ultralow-field NMR spectra.
Figure~\ref{Fig-Asymmetry}e schematically shows the basic interference effect resulting from the response of atomic magnetometer.
${\textrm{A}_\xi}(\nu ,{B_z})$ and ${\Phi_\xi}(\nu ,{B_z})$ ($\xi=x,y$) are, respectively, the amplitude and phase response functions.
Specifically,
the output signal of the atomic magnetometer $s(t)$ is an oscillating signal.
The amplitude of the oscillating $s(t)$ signal is
\begin{eqnarray}
 S_{tot}^2 &=& (S_x^2+S_y^2)(1+ \chi \cos \Delta \phi),
 \label{stot}
\end{eqnarray}
where $S_{x}= S_x(\nu ,{B_z}) = \alpha \textrm{A}_x (\nu ,{B_z}) B_{x0}$ and $S_y = S_y(\nu ,{B_z})= \alpha \textrm{A}_y (\nu ,{B_z}) B_{y0}$ respectively denote the amplitude of the output oscillating signal
when there $\textrm{only}$ exists the $x$- or $y$-component in the input magnetic field $\textbf{B}(t)=[B_{x0} \cos(2\pi \nu t+ \theta_{x0}), B_{y0} \cos(2\pi \nu t+ \theta_{y0}),B_z]$.
The proportionality constant $\alpha$ summarises, e.g. amplifier gains and conversion factors of detectors.
The term ``$\chi \cos \Delta \phi$'' represents the interference effect when the $x$- and $y$-components are both $\textrm{nonzero}$.
Here the interference phase $\Delta \phi=\Delta \Phi + (\theta_{x0}-\theta_{y0})$ and the interference contrast $ \chi \equiv 2S_x S_y/(S_x^2+S_y^2)$.
Based on this interference model, the asymmetric amplitudes of ZULF NMR spectra can be well explained~\cite{jianginterference}.
In addition,
a standard approach is presented to detecting and characterizing the interference effect in, but not limited to, atomic magnetometers,
enabling a higher confidence for gaining precise knowledge from NMR spectra~\cite{jianginterference}.


\section{Quantum control techniques}\label{control}

In a conventional high-field NMR experiment, the strong static magnetic
field ensures the wide separation of resonance frequencies for
non-zero-spin nuclei (for example, $^1$H, $^{13}$C, $^{15}$N, and $^{19}$F) allowing each to be
addressed selectively using AC magnetic field pulses \cite{vandersypen2005nmr}.
In contrast, as the resonance frequencies are not well separated (or even equal to zero) in ZULF NMR, the selection of individual spin species is
not straightforward \cite{tayler2016nuclear}, and thus
presents a challenge
and limits the possible applications of ZULF NMR \cite{jiang2018experimental}.
Nevertheless, quantum control methods are developed that could overcome the difficulties and achieve sophisticated manipulations of spins in ZULF $\textrm{NMR}$ \cite{bian2017universal,jiang2018numerical,ji2018time,tayler2016nuclear,sjolander2016transition,llor1995coherent,lee1987theory, llor1991scaling}.
These include zero-field spin echoes \cite{llor1991scaling}, rank-selective decoupling sequences \cite{llor1995coherent,llor1995coherent2}, universal quantum control based on composite pulses \cite{bian2017universal} and numerical optimization methods \cite{jiang2018numerical}, time-optimal control \cite{ji2018time}, and transition-selective pulses \cite{sjolander2016transition}.
Moreover, experiments based on these methods are performed~\cite{tayler2016nuclear,sjolander2016transition,jiang2018experimental,ji2018time}.
Experimental techniques to characterize the performance of quantum controls are also demonstrated~\cite{jiang2018experimental,ji2018time}.
In this section, we give a review of these quantum control methods and experiments in ZULF NMR.

\subsection{Controllability of ZULF NMR systems}

\begin{figure}[t]
	\centering
	\includegraphics[scale=1.1]{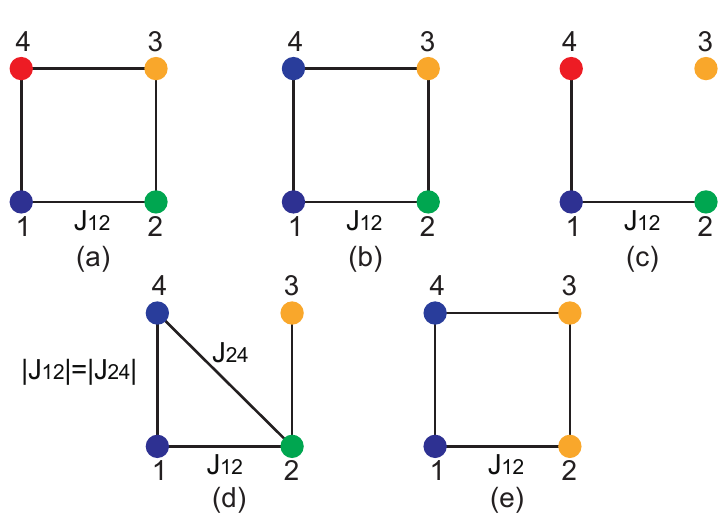}
	\caption{(Color online) Graphs representing the network of a $4$-spin system. Nuclear spins (colored balls and different colors representing different nuclear species) are labeled with different numbers. Graph (a) satisfies the controllability condition (i) and (b) satisfies condition (ii), so they can be completely controlled. Graphs (c), (d) and (e) satisfy neither condition (i) nor (ii). Reprinted with permission from ref. \cite{jiang2018numerical}, Copyright @ 2018 American Physical Society.}
	\label{bjfigconnectivity}
\end{figure}

The internal and control Hamiltonian of a liquid state NMR system at zero and ultralow field are given in Sec.~\ref{system}.
Before presenting how to realize quantum control in such systems, one important issue is whether, or under what conditions,
it is possible to control a quantum system using any physically permitted operations, i.e., the controllability problem.
In a closed quantum system, complete controllability refers to the ability of realizing any element in $\textrm{SU}(2^{n})$ \cite{dong2010quantum}.
The complete controllability of a spin system at zero and ultralow field can be described in terms of the graph \cite{albertini2002lie}:
the spin network is represented by a graph, as shown in Fig.~\ref{bjfigconnectivity}.
Each node representing a spin is labeled by a number; the edge connecting $i$th and $j$th spins corresponds to the J-coupling.
From ref.~\cite{albertini2002lie}, the complete controllability of the system is related to the connectivity of the graph.
Mathematically, a graph is connected when there is a path between every pair of nodes.
Then the above system is complete controllable if one of the following conditions is satisfied:

(i) The graph representing the spin system is connected, and gyromagnetic ratios of the spins are all different.

(ii) The graph representing the spin system is connected. The $m$ ($m \geq 1$) spins with unique $\gamma$ (each of them has different gyromagnetic ratio from all the others in the system) are defined in a set $S$. Define the rest spins with the same $\gamma$ to form sets $S_1$, $S_2$, ..., $S_p$, respectively. All the elements in $S_1$ to $S_p$ are transferred to $S$ by the following disintegration algorithm: 1. Define set $\Gamma=\{S_1, S_2, ..., S_p\}$. 2. If at least two spins $j$ and $k$ in $S_i$ and a spin $l$ in $S$ satisfy $J_{jl} \neq J_{kl}$, then transfer spin $j$ and $k$ in set $S_i$ to set $T$. If there is no element in $S$ and no set in $\Gamma$ satisfying this property, stop and put all the elements of set $T$ into set $S$. 3. Go back to step $2$ and repeat the process.

Note that condition (i) is a necessary and sufficient condition of complete controllability, while condition (ii) is a sufficient condition.

\subsection{Universal quantum control}\label{bjseccomposite}

\begin{figure*}[htb]
	\centering
	\includegraphics[scale=1.7]{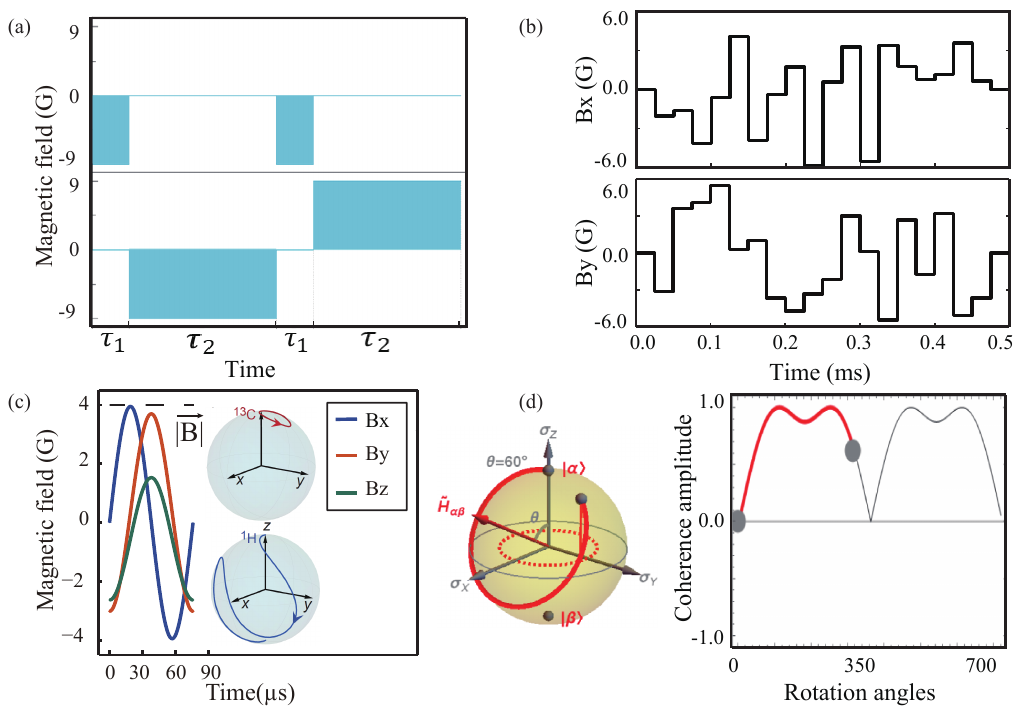}
	\caption{(Color online) Quantum control methods in ZULF NMR. (a) Composite pulse sequence realizing $\pi/2$ gate on $^{13}$C in a $^{13}$C-$^1$H system. Reprinted with permission from ref. \cite{bian2017universal}, Copyright @ 2017 American Physical Society. (b) Optimal control pulses for realizing $^{1}$H $\pi$ rotation along $z$ in $^{1}$H-$^{31}$P system. Reprinted with permission from ref. \cite{jiang2018numerical}, Copyright @ 2018 American Physical Society. (c) Time-optimal control fields and corresponding trajectories on the Bloch sphere for realizing $R^{H}_y(\pi)$.  Reprinted with permission from ref. \cite{ji2018time}, Copyright @ 2018 American Physical Society. (d) Bloch sphere representation showing how the density matrix evolves in the interaction frame during the application of the AC pulse \cite{sjolander2016transition}. Here the magnitude of the detected signal is proportional to the projection of the final state onto the $x$-$y$ plane. Reprinted with permission from ref.~\cite{sjolander2016transition}, Copyright @ 2016 American Chemical Society.}
	\label{bjcontrol4}
\end{figure*}

Heteronuclear spin systems satisfying condition (i) are completely controllable. This means that universal quantum control can be implemented in such systems \cite{nielsen2002quantum}.
Universal quantum control in ZULF NMR might find applications in several areas including correlation J-spectroscopy \cite{sjolander2020two}, chirality measurement \cite{king2017antisymmetric}, and Cosmic Axion Spin Precession Experiment (CASPEr) \cite{garcon2019constraints, budker2014proposal}.
A practical way to achieve universal control is to realize a set of universal logic gates, e.g., arbitrary single-qubit gates and two-qubit controlled-NOT gates (CNOT) \cite{nielsen2002quantum}.
Here a ``qubit'' is represented by a ``spin-$1/2$'' particle.
Implementing the above gates in heteronuclear spin systems satisfying condition (i) by composite pulses is studied in ref.~\cite{bian2017universal}.
Denote an arbitrary single-qubit gate on spin $i$
\begin{equation}
\label{single_gate}
U_{\mathbf{n}}^ i(\theta) = e^{-i \mathbf{n} \cdot \mathbf{I}_{i}  \theta},
\end{equation}
where $\mathbf{n}$ is the unit vector and $\theta$ is the rotation angle.
According to ref.~\cite{bian2017universal}, arbitrary single-qubit gates in such systems are realized by the following pulse sequence
\begin{equation}
\label{sin}
U_{\mathbf{n}}^1 (\theta) = U_{\mathbf{n}_{\perp}}^{2\sim n} (\pi)  e^{ -i H_{\textrm{dc}}(-{\mathbf{B_n}}) t/2}
U_{\mathbf{n}_{\perp}}^{2\sim n}{}^{\dagger } (\pi)  e^{ -i H_{\textrm{dc}}(-{\mathbf{B_n}}) t/2},
\end{equation}
where $\theta = \gamma_1 B t$ is the desired rotation angle, $|\mathbf{B_n}|=B$ is the control field strength, $U_{\mathbf{n}_{\perp}}^{2\sim n} (\pi)=U_{\mathbf{n}_{\perp}}^{2}(\pi)U_{\mathbf{n}_{\perp}}^{3}(\pi)...U_{\mathbf{n}_{\perp}}^{n}(\pi)$,
and
$H_{\textrm{dc}}(\textbf{B}) = - \sum\limits_j {{\gamma _j}{\textbf{I}_j} \cdot \textbf{B}}$.
${\mathbf{n}_{\perp}}$ denotes an arbitrary direction that is perpendicular to $\mathbf{n}$.
Spin $1$ is rotated by the desired angle, while the changes in other spins are refocused at the end of the sequence.
Note that J-couplings are omitted during DC pulses.
A schematic diagram of implementing the above method is demonstrated in Fig.~\ref{bjcontrol4}a.
The sequence relies on single-qubit $\pi$ pulses on rest of the qubits. 
Given that the gyromagnetic ratios of different qubits are different, one could construct the $\pi$ pulses 
through applying a constant control field along the desired axis for some fine-tuned time $\tau$.  
This leads to high fidelity $\pi$ pulses in typical small spin systems \cite{bian2017universal}.
Here the fidelity is defined by
\begin{equation}
F=|\textrm{Tr}(U^{\dagger}_{\textrm{ideal}}U)|/2^n,
\end{equation}
which describes the accuracy of a realized unitary operation $U$ with respect to the ideal one $U_{\textrm{id}}$.
For example, in a two-qubit $^{13}$C-$^{1}$H system, 
as $\gamma_\textrm{H}/\gamma_\textrm{C} \approx 4$, one could choose $\tau=\pi/(B\gamma_C)$ such that $^{13}$C undergoes 
$\pi$ rotation, while $^{1}$H undergoes 
approximately $4\pi$ rotation, and a $\pi$ pulse on $^{13}$C is thus realized with $F \approx 0.9994$. In general, $\tau$ 
could be found numerically through maximizing $F$ in a given range of pulse duration. For example, 
in a C-H-F system,
$\pi$ pulse on both $^{1}$H and $^{19}$F simultaneously (leaving $13$C alone) is obtained with $F=0.9933$ through 
the numerical method.
For complicated spin systems, the above method may fail \cite{bian2017universal}.
In that case, one could apply numerical optimal control method as explained in Sec. \ref{biansecgrape}.

One still need CNOT gates between arbitrary spin $i$ and $j$ to achieve universal control. Its matrix form in $I_z$ basis $\{ |0 \rangle_i | 0 \rangle_j ,  |0 \rangle_i | 1 \rangle_j,  |1 \rangle_i | 0 \rangle_j,  |1 \rangle_i | 1 \rangle_j\}$ with $I_z |0 \rangle = \frac{1}{2}|0 \rangle $ and  $I_z |1 \rangle = - \frac{1}{2}|1 \rangle $ reads:
\begin{equation}
\label{ocnot}
\textrm{C}\textrm{NOT}_{ij} = \left( {\begin{array}{*{20}{c}}
	1&0&0&0\\
	0&1&0&0\\
	0&0&0&1\\
	0&0&1&0
	\end{array}} \right),
\end{equation}
where spin $i$ is the control spin, and spin $j$ is the target spin. This operation flips spin $j$ (target spin) when spin $i$ (control spin) is in state $|1\rangle$ and does nothing when spin $i$ is in state $|0\rangle$. In a two-qubit system, this operation could be further decomposed into \cite{vandersypen2005nmr}
\begin{equation}
\label{bjeqcnot}
\textrm{C}\textrm{NOT}_{ij} =\sqrt{i} U_z^i (\frac{\pi}{2})  U_z^j{}^{\dagger } (\frac{\pi}{2})  U_x^j (\frac{\pi}{2})
U^{(i,j)}_{\text{zz}} (\frac{\pi}{2})  U_y^j (\frac{\pi}{2}),
\end{equation}
where
\begin{equation}
U^{(i,j)}_{\text{zz}} (\phi)=e^{-iH^{(i,j)}_0t}U_z^j(\pi)e^{-iH^{(i,j)}_0t}U_z^j{}^{\dagger }(\pi),
\label{Uzz}
\end{equation}
$H^{(i,j)}_0= {{2 \pi J_{ij}}}  \mathbf{I}_i \cdot \mathbf{I}_j$ and $\phi \equiv 2 \pi J_{ij}t$ ($J_{ij}>0$).
For spin systems with $n$ spins ($n>2$), the main barrier to implementing the CNOT gate is the implementation of $U^{(i,j)}_{\text{zz}} (\theta)$ in a large coupled spin network, where only the coupling $H^{(i,j)}_0$ is active.
To achieve this, one needs to turn off the undesired couplings, as achieved by refocusing \cite{vandersypen2005nmr} in high-field NMR.
This can be achieved in zero-field NMR via a concatenated scheme by recursively building on a basic sequence \cite{bian2017universal}.

The above universal quantum control scheme was experimentally demonstrated in ref.~\cite{jiang2018experimental}.
Single-qubit control for $^{13}$C and $^{1}$H and two-qubit control via a $\textrm{CNOT}$ gate in $^{13}$C-formic acid was realized on a home-built ZULF $\textrm{NMR}$ spectrometer~\cite{jiang2018experimental}.
The experimental apparatus and procedure were outlined in Sec.~\ref{subsecexp}.
The average fidelity for $^{13}$C single-qubit control was $f_{\textrm{avg}}=0.9960(2)$.
This was obtained by a quantum information-inspired randomized benchmarking method that will be explained in Sec.~\ref{bjchaprb}.
In this experiment, it was the unitary errors which originated from pulse imperfections that principally limited the single-spin control fidelity.
The CNOT$_{\textrm{HC}}$ gate was also realized in ref.~\cite{jiang2018experimental}, and the fidelity was around $F=0.9877$.


\subsection{Numerical optimal control}\label{biansecgrape}

In ref.~\cite{jiang2018numerical}, numerical optimization method based on the gradient ascent pulse engineering (GRAPE) algorithm~\cite{khaneja2005optimal} is applied to manipulate nuclear spins in zero-field NMR.
In particular, it is shown that this method is capable of designing control pulses for homonuclear spin systems satisfying condition (ii).
This is not achievable via the composite pulse scheme presented in Sec. \ref{bjseccomposite}.
Thus the numerical optimization method greatly improves the ability to manipulate spin systems at zero field~\cite{jiang2018numerical}.

GRAPE is a gradient-based numerical optimization method \cite{khaneja2005optimal}. It promises to achieve precise and robust quantum control in a wide range of quantum systems, and is widely used in the magnetic resonance community and quantum information science community \cite{glaser2015training,tovsner2009optimal}. As an example, $\pi$ pulse on $^1$H in $^1$H-$^{31}$P system (i.e., Phosphorous acid) is obtained by the GRAPE method \cite{jiang2018numerical}.
The optimal control fields are illustrated in Fig.~\ref{bjcontrol4}b.
The total duration $T=0.5$ ms (i.e., number of pieces:
$20$ and piecewise duration: $25$ $\mu$s) and the fidelity $F=0.9997$.
It was further demonstrated in ref.~\cite{jiang2018numerical} that universal control could also be achieved by this method.
In particular, it is generally considered to be formidable for manipulating homonuclear spins at zero field due to the same gyromagnetic ratios.
However,
homonuclear spins in systems satisfying condition (ii) in Sec. \ref{bjseccomposite}  can be distinguished utilizing the information on different J-couplings.
For instance, in $^1$H-$^{13}$C$_1$-$^{13}$C$_2$ system (i.e., Trichloroethylene) composed of two $^{13}$C spins,
the system has different J-couplings between $^{13}$C and $^1$H,
i.e., J$_{\text{H}\text{C}_1}=202.3$ Hz, J$_{\text{H}\text{C}_2}=13.4$ Hz.
This system satisfies condition (ii) and hence it is universally controllable.
By using the GRAPE method,
a $\pi$ $z$-pulse on $^{13}$C$_1$ is achieved with $T=90$ ms (i.e., number of pieces: 450 and piecewise duration: 200 $\mu\text{s}$) and $F=0.9999$.

Moreover, the requirement of robustness against pulse imperfections could be incorporated into the GRAPE method. According to Sec.~\ref{bjseccomposite},
the main errors come from pulse imperfections that principally limit the performance of control.
In ref.~\cite{jiang2018numerical}, the robustness tests (against pulse imperfections)  for single-qubit and two-qubit gates in $^1$H-$^{31}$P system are demonstrated.
The result indicates that the optimal control pulses are more robust compared with the composite pulse scheme presented in Sec.~\ref{bjseccomposite}.
Therefore,
the GRAPE method allows for decreasing requirement of hardware stability and calibration accuracy in practical experiments. Note that this method could be straightforwardly generalized to ultralow-field NMR. Together with its high fidelity and capability of designing control fields in homonuclear spin systems, future experiments based on this method is worth exploring.

\subsection{Time optimal control}

Time-optimal control (TOC) problems in quantum systems are ubiquitous and important in multiple applications \cite{glaser2015training,khaneja2001time,carlini2006time,d2007introduction}. Because the inevitable noise from the environment degrades quantum states and operations over time, inducing quantum dynamics in minimal time utilizing TOC becomes a preferable choice \cite{geng2016}. The analytic knowledge of the TOC is useful even in cases where such a control is not the one physically implemented. It gives information about the inherent time limitations of the system, therefore indicating a benchmark for the time of any control law \cite{ji2018time}. However, analytic solutions are
rare in optimal control and theoretical calculations have to be complemented by difficult numerical simulations \cite{glaser2015training}.
Previous works mainly considered time optimization with controls which address spin individually
\cite{khaneja2001time,geng2016}. However, this is not the case in zero-field NMR,
where all spins are simultaneously affected by the control fields.
In ref.~\cite{ji2018time}, TOC for arbitrary single-qubit rotation in a two-qubit system under simultaneous control is analytically obtained and experimentally implemented in a zero-field NMR experiment. This further improves the ability to manipulate spins in ZULF NMR.

To obtain the TOC, the following procedure is adopted in \cite{ji2018time}.
The Pontryagin's maximum principle (PMP) was first applied to reduce the number of unknown parameters to $7$.
Then, a newly developed symmetry reduction technique is applied to further reduce the number of unknown parameters to $3$.
Finally, the TOC and optimal time are obtained by solving a set of equations.
For example, the control magnetic fields $\mathbf{B_n}=(B_x,B_y,B_z)$ are analytically obtained to realize $R_{y}^ H(\theta)$ in a two-qubit $^{1}$H-$^{13}$C system,
as illustrated in Fig.~\ref{bjcontrol4}c.
There is a $70\% \sim 80\%$ reduction in the time cost of $R_{y}^ H(\theta)$ compared to the composite pulse scheme introduced in Sec.~\ref{bjseccomposite}.
The experiments were performed in the $^{13}$C-formic acid sample using a home-built zero-field NMR spectrometer.
The average fidelity for $^1$H single-qubit TOC control was $f_{avg}=0.99$~\cite{ji2018time}.

\subsection{Transition-selective pulses}

Apart from the above universal control methods, there are alternative methods adding an increased capability for spin control in ZULF NMR \cite{sjolander2016transition,tayler2016nuclear}.
For example, the use of high-field selectivity in ZULF NMR was reported in ref.~\cite{tayler2016nuclear},
where a magnetic field was temporarily imposed on the sample,
allowing one to apply AC pulses that individually address different spin species.
Moreover, many NMR techniques rely on selectivity beyond differentiating spin species, to the extent of addressing individual transitions~\cite{freeman1991selective}.
In ref.~\cite{sjolander2016transition},
frequency selective pulses in ultralow-field regime were demonstrated with a typical excitation bandwidth of $0.5$-$5$ Hz.
The resonant AC pulses were utilized to drive spin populations directly between the zero-field eigenstates and therefore allowed greater control over which transitions are excited.
In ref.~\cite{sjolander2016transition}, the evolution of the system under the AC magnetic field, J-coupling and a small static field were obtained analytically in an interaction frame.
This is illustrated in Fig.~\ref{bjcontrol4}d.
The result is that the total Hamiltonian can be block diagonalized to a time-independent Hamiltonian acting on a two level system.

\begin{figure*}[t]  
	\makeatletter
	\def\@captype{figure}
	\makeatother
	\includegraphics[scale=1.18]{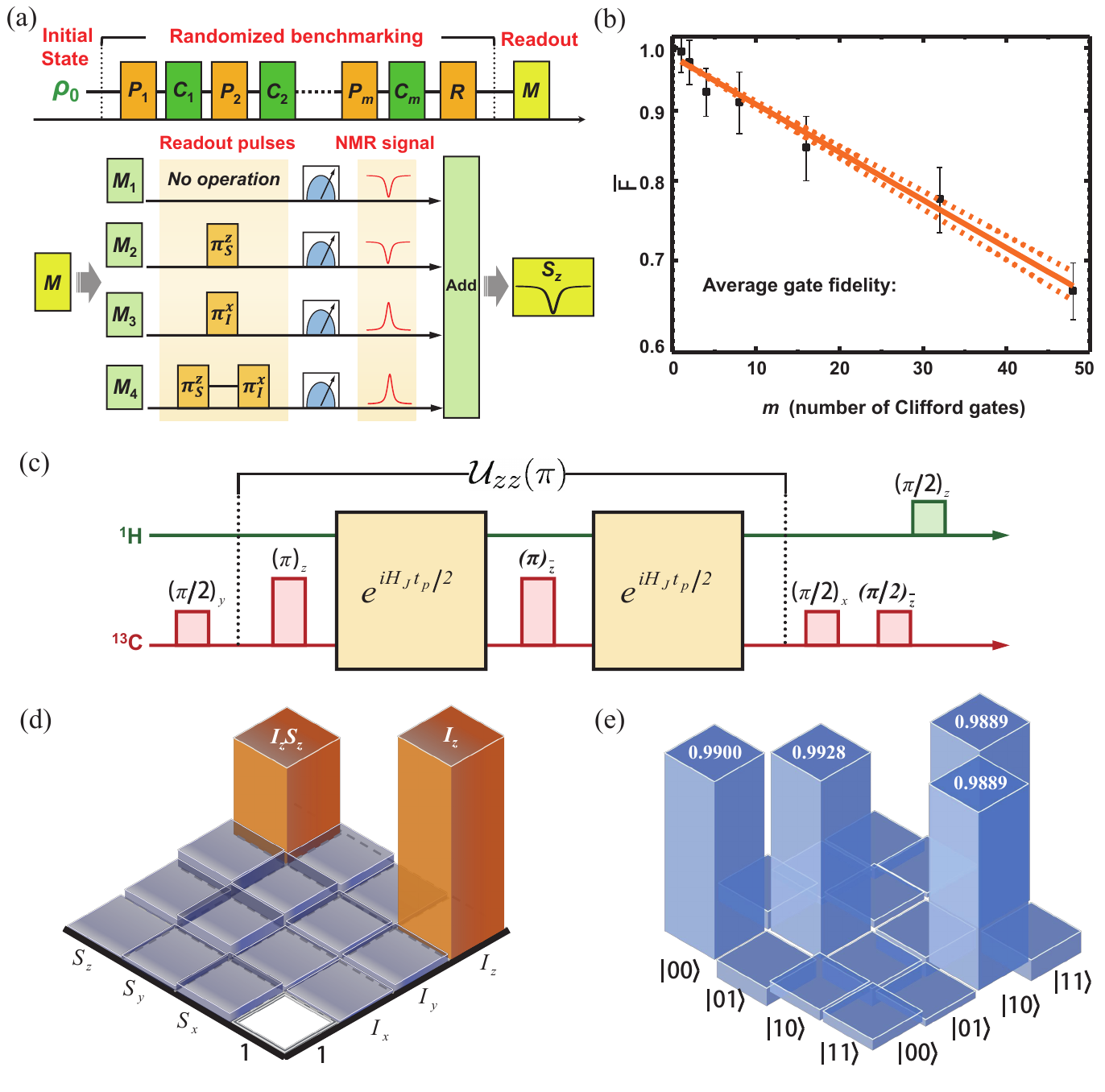}
	\caption{(Color online) (a) Clifford-based randomized benchmarking. The signals acquired using four different independent pulse
		sequences (readout operations), $M \in \{\textrm{No operation}, \pi^{z}_S, \pi^{x}_{I},\pi^{z}_S-\pi^{x}_{I}\}$, are
		averaged together. Random sequences with $P=e^{\pm i\pi A}$ and $C=e^{\pm i(\pi/2) B}$
		are applied for each sequence length $m$, where the
		Clifford gates are realized by combined operations $PC$. Here, $A \in \{\mathbf{1},S_x,S_y,S_z\}$ and $B \in \{S_x,S_y,S_z\}$. The recovery operation $R$ is chosen to
		return the system to the initial state in the absence of control error. (b)  Randomized benchmarking results for $^{13}$C single-spin control. Each point is an average over $32$ random sequences of $m$ Clifford gates, and the error bars indicate the standard error of the mean. A single exponential decay shown with a solid line is used to fit the fidelity decay. The randomized benchmarking results yield an average
		error per Clifford gate $\epsilon_g = 0.0040(2)$, and an imperfection of the state
		initialization and readout $d_{if} = 0.0141$.    (c) Pulse sequences for implementing the CNOT gate, as explained in Sec. \ref{bjseccomposite}. (d) Output of the CNOT gate applied to the sudden state. (e) Reconstructed CNOT gate in the computational basis. The fidelity of the CNOT gate is $0.9877 (2)$. Reprinted with permission from ref. \cite{jiang2018experimental}, Copyright @ 2018 American Association for the Advancement of Science.}
	\label{bjfigrb}
\end{figure*}

Using this method, one is able to discriminate between signals belonging to manifolds of different total proton angular momentum in the zero-field spectrum of [$^{15}$N, $^{13}$C$_2$]acetonitrile.
It is further shown that the sense of rotation of pulsed fields may select between positive and negative changes in angular-momentum projection.
Using two orthogonal pulsing coils each generating an AC field, a rotating magnetic field is generated.
It allows one to select transitions with $\Delta m_F=+1$ or $\Delta m_F=-1$.
A weak DC field is applied to the $^{13}$C-formic acid, and the resulting spectrum contains two observable transitions centered about the J-coupling frequency.
By applying the above rotating magnetic field, both transitions could be addressed selectively,
sidestepping the usual limit on frequency selectivity. These techniques should facilitate ZULF NMR spectroscopy of larger,
more demanding spin systems or mixtures and open a way to adapting a suit of established high-field experiments to ultralow field~\cite{sjolander2016transition,tayler2016nuclear}.

\subsection{Evaluation of quantum control}\label{bjchaprb}

Being able to quantify the performance of coherent quantum control is important for the development of quantum architectures~\cite{vandersypen2005nmr}.
Full characterization of any quantum process, and then calculation of the fidelity of control, is possible through a procedure known as quantum process tomography ($\textrm{QPT}$) \cite{vandersypen2005nmr}.
However, QPT requires an exponential number of experiments, making it experimentally prohibitive for quantum systems larger than a few qubits.
To overcome this, in ref.~\cite{knill2008randomized}, it was reported the randomized benchmarking (RB) method to determine the error probability per gate in computational contexts.
In a recent work~\cite{jiang2018experimental}, Clifford-based RB was applied to estimate the fidelity of single-qubit controls in the $^{13}$C-$^{1}$H system in ZULF NMR.
The randomized benchmarking pulse sequences are shown in Fig.~\ref{bjfigrb}a. The sudden state is selected as the initial state.
 To measure the coefficient of $S_z$ independently,
the same temporal averaging method as the state tomography is adopted,
as explained in Fig.~\ref{bjfigrb}a.
 Measuring the decay of the coefficient of $S_z$ with respect to the number
 ($m$) of randomized Clifford gates in the benchmarking sequence
 yields the average fidelity for $^{13}$C single-spin control. By averaging
 the coefficients of $S_z$ over $k$ different randomized benchmarking sequences
 with the same length $m$, and normalizing this averaged value
 to that of $m=0$, the normalized signal $\bar{F}$ can be written as  $\bar{F}=(1-d_{if})(1-2\epsilon_g)^{m}$,
 where $d_{if}$ is due to the imperfection of the state initialization
 and readout, and $\epsilon_g$ is the average error per Clifford gate \cite{jiang2018experimental}. As explained
 in Fig.~\ref{bjfigrb}b, the average fidelity for $^{13}$C
 single-spin control is estimated to be $f_{\textrm{avg}} = 1 - 0.0040(2) = 0.9960(2)$, which is resilient
 to the state preparation and measurement errors.

 \begin{figure*}[htb]  
	\makeatletter
\centering
	\def\@captype{figure}
	\makeatother
	\includegraphics[scale=1.2]{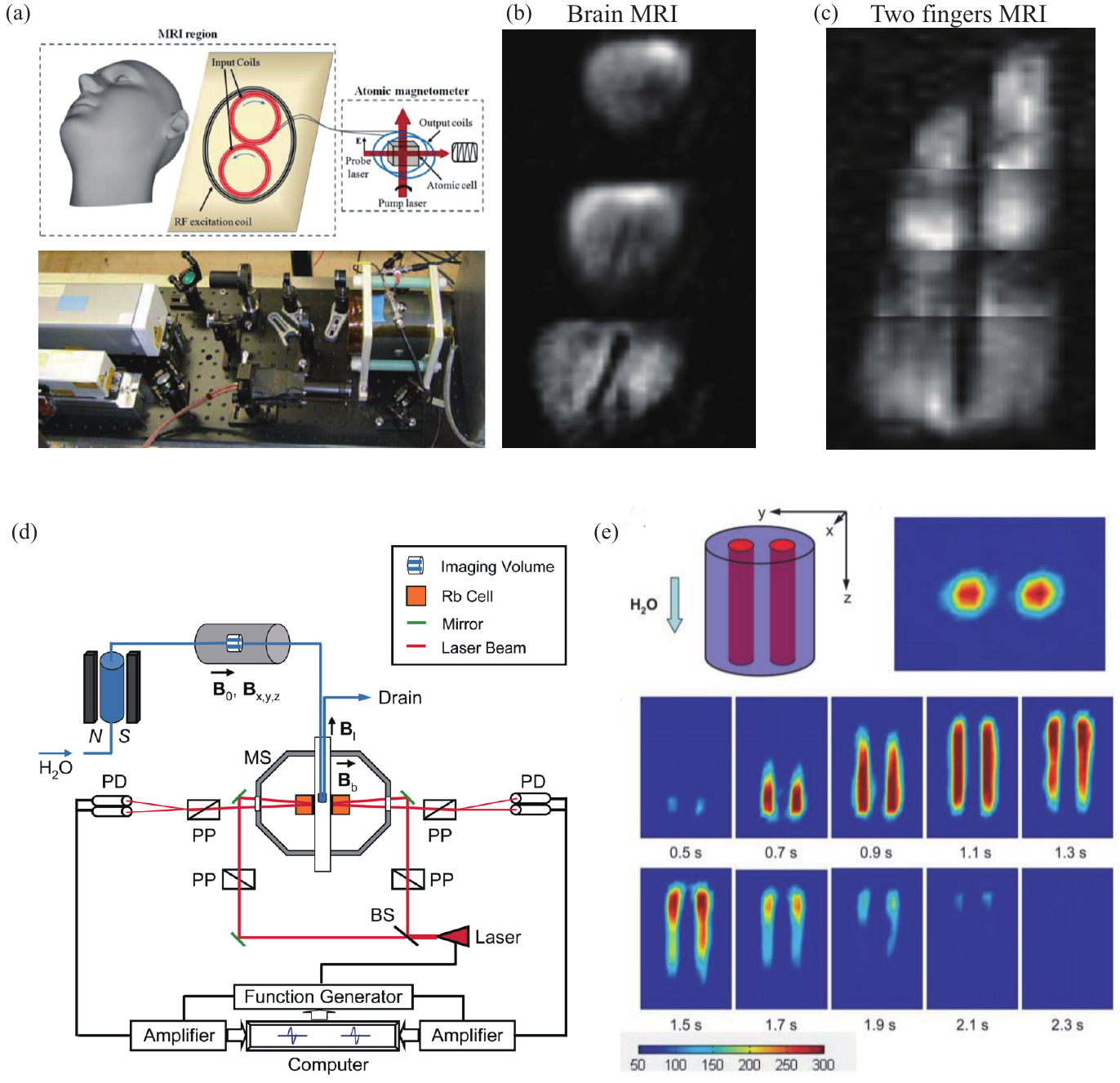}
	\caption{(Color online) (a) Ultralow-field MRI detection apparatus. (b) Ultralow-field MRI of human brain. These images reveal some obvious brain features and boundaries. Reprinted with permission from ref.~\cite{savukov2013magnetic}, Copyright @ 2013 American Chemical Society. (c) Ultralow-field MRI of two fingers. Reprinted with permission from ref.~\cite{savukov2013anatomical}, Copyright @ 2013 Elsevier. (d) and (e) Schematic of the experimental setup and MRI of two parallel cylindrical channels. Reprinted with permission from ref.~\cite{xu2006magnetic}, Copyright @ 2006 National Academy of Sciences.}
	\label{mri}
\end{figure*}

 \begin{figure*}[t]  
	\makeatletter
\centering
	\def\@captype{figure}
	\makeatother
	\includegraphics[scale=1.7]{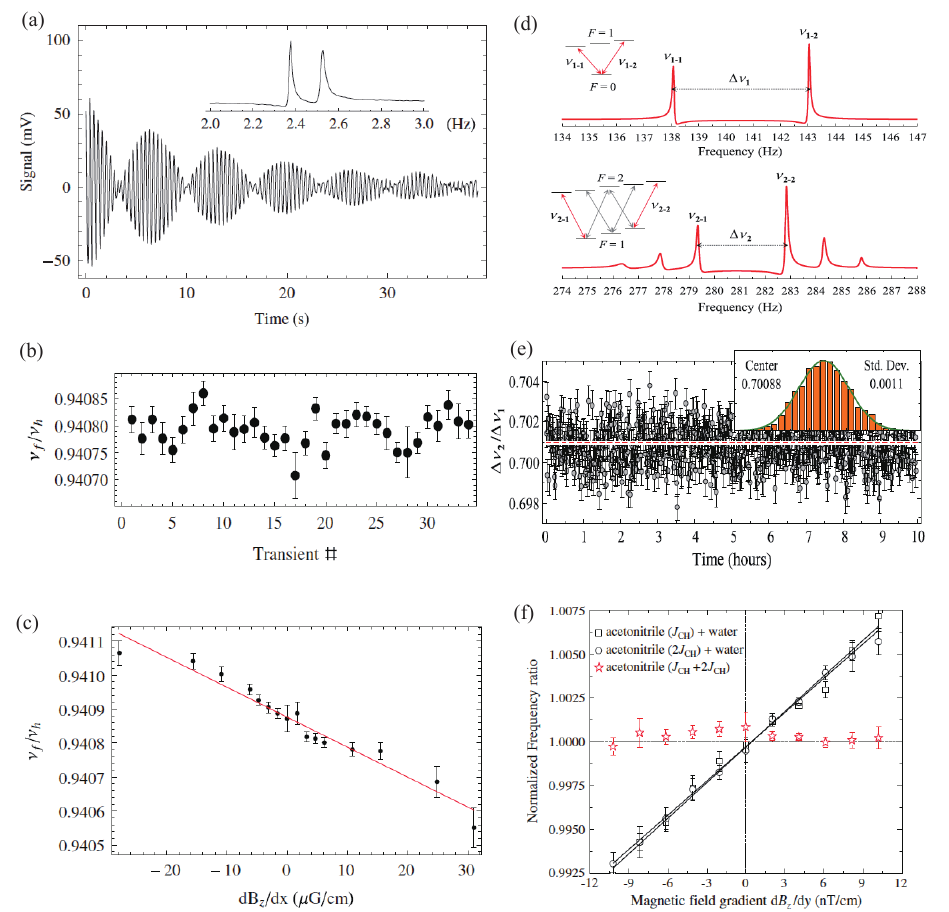}
	\caption{(Color online) (a) Pentane-hexafluorobenzene comagnetometer based on ZULF NMR. Ultralow-field NMR signals from a mixture of $^1$H-pentane (C$_5$H$_{12}$) and $^{19}$F-hexafluorobenzene (C$_6$F$_6$). The time-domain signal is well described by a sum of two exponentially decaying sinusoids. As shown in the inset, the spectrum shows two peaks at the Larmor precession frequencies of $^1$H and $^{19}$F nuclei ($\nu_h$ and $\nu_f$). $\nu_h$ and $\nu_f$ are both proportional to external magnetic field, but $\nu_f/\nu_h$ is insensitive to magnetic field and can be used to construct as a comagnetometer. (b) The ratio $\nu_f/\nu_h$ versus external magnetic field modulation. The modulation on the ratio is not visible, which is in good agreement with theoretical prediction. (c) Effects of magnetic field gradients on the pentane-hexafluorobenzene comagnetometer. Reprinted with permission from ref.~\cite{RNLiquidstate}, Copyright @ 2012 American Physical Society. (d) Comagnetometer based on a liquid of identical molecules. Experimentally measured J-coupling spectrum of acetonitrile-$2$-$^{13}$C ($^{13}$CH$_3$CN) at $J_{CH}$ and 2$J_{CH}$ in a 100 nT bias field along $z$. The splittings $\Delta \nu_1$ and $\Delta \nu_2$ are both proportional to external magnetic field, and can be employed to realize a comagnetometer based on an ensemble of identical molecules. (e) The ratio $\Delta \nu_2/\Delta \nu_1$ remains constant although an oscillating magnetic field is applied. (f) In the presence of a constant bias magnetic field $B_z=100$~nT, the normalized frequency ratio as a function of the gradient $dB_z/dy$. Reprinted with permission from ref.~\cite{RNComagnetometer}, Copyright @ 2018 American Physical Society.}
	\label{comagnetometer}
\end{figure*}

 The performance of the CNOT gate in  the $^{13}$C (target spin)-$^{1}$H (control spin) system is also evaluated in ref.~\cite{jiang2018experimental}.
 The CNOT gate is implemented according to the method presented in Fig.~\ref{bjfigrb}c and Sec. \ref{bjseccomposite}.
 As the experimental error is dominated by unitary error,
 a partial quantum process tomography (see ref.~\cite{jiang2018experimental}) is implemented to characterize the CNOT gate.
 It requires fewer measurements than standard QPT. It takes two independent initial states as the input states and measure the corresponding output states after applying the CNOT gate.
 The $\textrm{CNOT}$ gate is then reconstructed by using a numerical minimization technique \cite{jiang2018experimental}.
 One of the output state is illustrated in Fig.~\ref{bjfigrb}d.
 The reconstructed CNOT gate is illustrated in Fig.~\ref{bjfigrb}e, and then the fidelity is calculated to be $0.9877(2)$.

\section{Imaging}

MRI is a widely used non-invasive technique for materials research and medical diagnosis~\cite{liang2000principles}.
Because the image quality improves with the magnetic field strength, conventional clinical scanners use fields as high as 1-3 T.
However, traditional MRI instruments are immobile and expensive, in particular they have long been off-limits to patients with pacemakers.
Also, high-field MRI is not applicable to metallic materials because of the short penetration depth of radiofrequency radiation.
To overcome the difficulties mentioned above,
one solution is to perform MRI at ultralow field on the order of microtesla~\cite{xu2006magnetic, savukov2013magnetic, savukov2013anatomical}.
In particular, ultralow-field MRI instruments can be constructed at much lower cost, can be more portable, and
can have many other advantages including higher contrast to anomalies and absence of susceptibility artifacts.
Starting with the early work with SQUIDs as the low-field MRI detector~\cite{zotev2008microtesla},
atomic magnetometers have surpassed the sensitivity of SQUIDs and could be a promising detector for MRI~\cite{allred2002high, dang2010ultrahigh}.
References~\cite{savukov2013magnetic, savukov2013anatomical} reported ultralow-field MRI systems for human brain and fingers, as shown in Fig.~\ref{mri}a, b and c.
In this MRI system, a $\sim$80 mT prepolarization field first prepolarizes spins and then is switched off before the MRI sequence begins,
and then spins evolve and are detected under 4 mT field through a pick-up coil transferring to an atomic magnetometer.
Moreover, it is also possible to perform MRI of metallic materials, which are of extensive interest because metallic materials broadly used in filtration, catalysis, and biomedicine.
Xu et al.~\cite{xu2006magnetic} reported remote-detection imaging (see Fig.~\ref{mri}d, e) where water was first prepolarized through a permanent magnet and flowed into a spatially separated encoding region with 31 G bias field,
and detection was performed with an atomic magnetometer at ultralow field.

Although the demonstrated MRI image quality is much lower than that typical in high-field MRI,
there are still potential to improve the current MRI quality such as improving the prepolarization field and optimizing the sensitivity of atomic magnetometers.
The ultralow-field MRI can find practical applications where high-field MRI is restricted.
For example, the ultralow-field MRI removes the off-limits to patients with pacemakers and provides a safe way of imaging for such patients.
In addition, such a MRI system could be very portable and thus can quickly move to patients without the need of relocating them to MRI facility.
This should be particular important because the relocation is dangerous, especially when life support system needs to be connected~\cite{savukov2013magnetic}.

\section{NMR-based quantum devices}

ZULF NMR operates at zero or ultralow fields,
which is particularly appropriate for precise measurements without electromagnetic interference and operate as sensitive quantum devices.
In this section, we review recent developments of comagnetometer and Floquet maser based on ZULF NMR.

\subsection{Comagnetometer}

Liquid-state comagnetometers based on ZULF NMR were demonstrated in refs.~\cite{RNLiquidstate,RNComagnetometer}.
As shown in Fig.~\ref{comagnetometer}a,
the comagnetometer is based on ultralow-field NMR of binary mixtures of mutually miscible solvents, for example a mixture of pentane and hexafluorobenzene,
which is similar to that of noble gas comagnetometer.
As shown in the inset,
the spectrum demonstrated two peaks at the Larmor precession frequencies of $^1$H and $^{19}$F nuclei ($\nu_h$ and $\nu_f$, respectively).
Because $\nu_h$ and $\nu_f$ are both proportional to external magnetic field, $\nu_f/\nu_h$ is insensitive on external magnetic fields (Fig.~\ref{comagnetometer}b).
Therefore, they can be employed to realize a comagnetometer,
which can resist magnetic field interference but retains sensitivity to Zeeman-like nonmagnetic spin interactions~\cite{wu2019search, garcon2019constraints}.
In such a comagnetometer, however, magnetic field gradients are one of the major sources of systematic errors (see Fig.~\ref{comagnetometer}c)
because there exists some spatial separation between the ensemble-averaged position of different spin species.

To overcome this difficulty, reference~\cite{RNComagnetometer} reported a single species comagnetometer,
in which different nuclear spins are probed within the same molecules.
Specifically, under the influence of a bias magnetic field, the J-coupling resonance lines at different frequencies split into separate peaks, as shown in Fig.~\ref{comagnetometer}d.
The frequency separation between the split peaks for each J-coupling resonance has distinct linear coefficients with respect to the magnetic field,
i.e., $\Delta \nu_1=(\gamma_h + \gamma_c)B_z$ and $\Delta \nu_2=\frac{1}{2}(\gamma_h + 3 \gamma_c)B_z$.
As shown in Fig.~\ref{comagnetometer}e, the ratio $\Delta \nu_2/ \Delta \nu_1$ is insensitive to external magnetic field changes.
Thus, measurements of these splittings can be employed as a comagnetometer.
In particular, because the probed spins in a same molecule interact with same magnetic field,
such a comagnetometer can greatly resist the perturbation of magnetic field gradients (Fig.~\ref{comagnetometer}f).

Liquid-state comagnetometers based on ZULF NMR are suitable for precision measurements,
such as sensitive gyroscopes and to search for spin-gravity, permanent electric dipole moments, axionlike dark matter and other fields in fundamental physics~\cite{wu2019search, garcon2019constraints}.

\subsection{Floquet maser}

\begin{figure}[t]  
	\makeatletter
\centering
	\def\@captype{figure}
	\makeatother
	\includegraphics[scale=0.15]{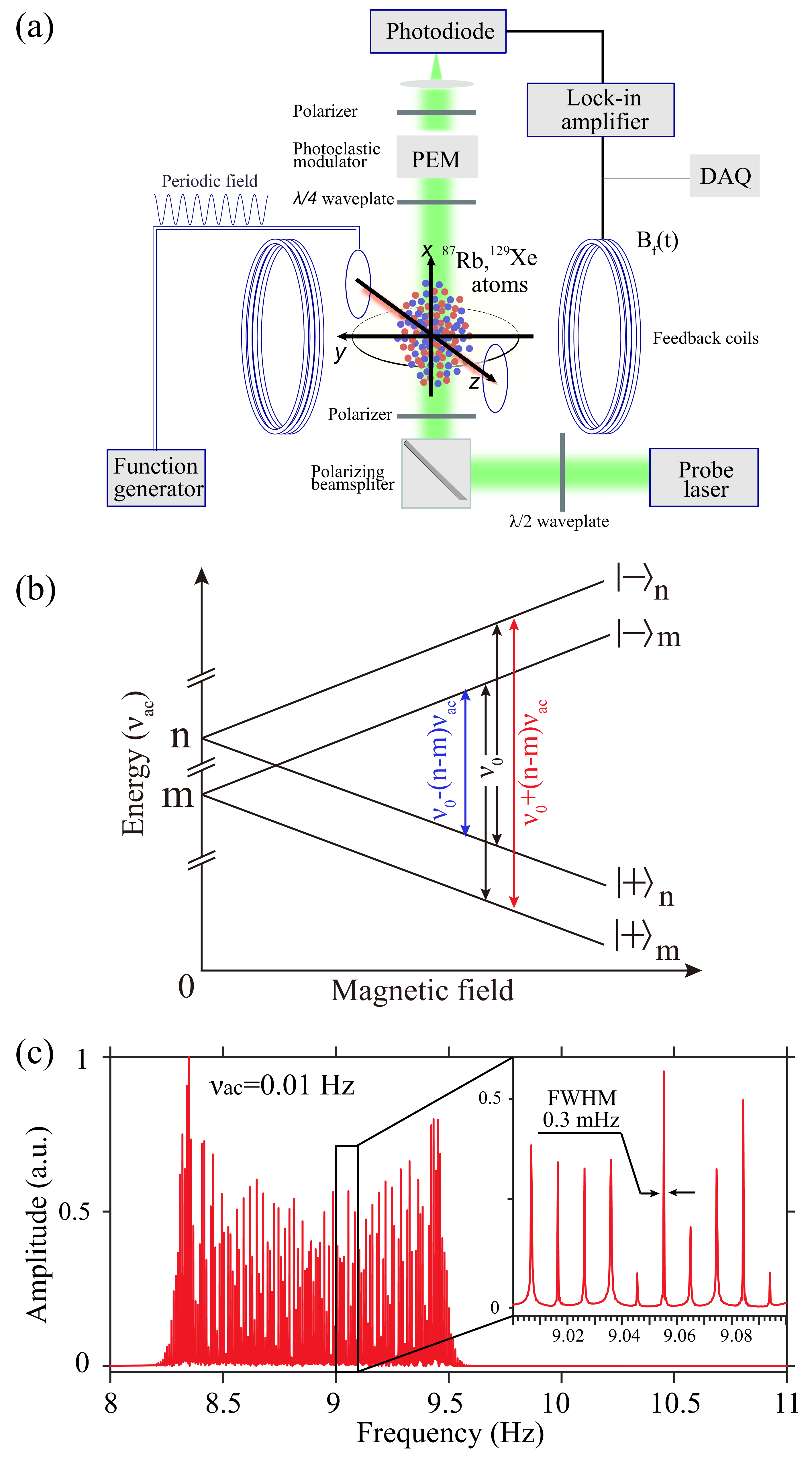}
	\caption{(Color online) (a) Schematic of the $^{129}$Xe nuclear-spin maser. The $^{129}$Xe nuclei are polarized and detected via spin-exchange collisions with optically pumped $^{87}$Rb. (b) Floquet states of a periodically driven Floquet system. (c) Spectrum based on Floquet $^{129}$Xe nuclear-spin maser. Reprinted from ref.~\cite{jiang2019floquet}.}
	\label{maser}
\end{figure}

The invention of the maser stimulated many revolutionary technologies such as lasers and atomic clocks~\cite{gordon1955maser, oxborrow2012room}.
Despite enormous progress,
the realizations of masers are still confined to a limited variety of systems,
in particular, the physics of masers remains unexplored in periodically driven (Floquet) $\textrm{systems}$,
which are generally defined by time-periodic Hamiltonians and enable to observe many exotic phenomena such as time crystals.
Recently we reported the first theoretical and experimental demonstration of a Floquet based maser~\cite{jiang2019floquet} comprised of periodically driven $^{129}$Xe spins in an vapor cell.
In contrast to liquid-state ZULF NMR,
We used noble gas atoms $^{129}$Xe with nuclear spin $I=1/2$ in a setup depicted in Fig.~\ref{maser}a.
A 0.5~cm$^3$ cubic vapor cell made from pyrex glass contains 5~torr $^{129}$Xe, 250~torr N$_2$, and a droplet of enriched $^{87}$Rb.
$^{129}$Xe spins are polarized by spin-exchange collisions with optically-pumped $^{87}$Rb atoms in a bias magnetic field $B_0$ ($\approx 750$~nT) along the polarized direction ($z$ axis).
Similar to a microwave cavity in conventional masers~\cite{gordon1955maser, oxborrow2012room},
the $^{129}$Xe spins in experiments are embedded in a feedback circuit, as shown in Fig.~\ref{maser}a,
which employs an atomic magnetometer as a sensitive detector of $^{129}$Xe spins and
simultaneously supplies the real-time output audio-frequency signal of the magnetometer to the spins.

Unlike the common masers that exploit inherent transitions,
the realized maser called "Floquet maser" oscillates at the frequencies of transitions between Floquet states shown in Fig.~\ref{maser}b.
Using the Floquet maser technique,
ultrahigh-resolution spectra of the Floquet system are observed with two orders of magnitude better resolution compared to that limited by decoherence (Fig.~\ref{maser}c).
As the spectral resolution is greatly increased,
a different regime emerges where high-order Floquet sidebands become significant and complex spectra are expected,
enabling accurate measurement of physical parameters,
e.g., atomic scalar and tensor polarizabilities, magnetic fields, ultralight bosonic exotic fields, and multiphoton coherences.
In particular, the Floquet maser could be extremely sensitive detector of low-frequency magnetic fields,
which is significantly better than state-of-the-art magnetometers.

\section{Tests of fundamental physics}

The applications of ZULF NMR are not limited to applied science,
but also can be applicable to tests of fundamental physics.
In particular, ZULF NMR-based quantum devices have extensively studied to measure magnetic fields or pseudo-magnetic fields with high sensitivity,
which are well suited to test of fundamental physics including the searches for spin-gravity coupling~\cite{RNComagnetometer, RNLiquidstate},
spin-axion coupling~\cite{garcon2019constraints, wu2019search, jiang2019floquet}, molecular parity nonconservation~\cite{king2017antisymmetric}.
The proposal of Cosmic Axion Spin Precession Experiment proposed the use of NMR techniques to search for dark-matter-driven spin precession~\cite{budker2014proposal}.
Based on this, the potential couplings between nuclear spins and axions are similar to the Zeeman interactions,
enabling their direct detection via high-resolution ZULF NMR spectra.
Reference~\cite{garcon2019constraints} reported
that as nuclear spins move through the galactic dark-matter halo,
they couple to dark matter and behave as if they were in an oscillating magnetic field, generating a dark-matter-driven NMR signal.
No dark matter signal was detected above background, establishing
new experimental bounds for dark matter bosons with masses ranging from $1.8\times 10^{-16}$ to $7.8\times 10^{-14}$~eV
(corresponding to Compton frequencies ranging from $\sim$45 mHz to 19 Hz).
To explore the window of lower Compton frequencies,
low-frequency magnetic field may cause electromagnetic interference because ZULF NMR spectra are also sensitive to real magnetic fields.
To overcome this, reference~\cite{wu2019search} reported the search of ultralight axion dark matter via a liquid-state comagnetoemter,
ranging from $10^{-22}$ to $1.3\times 10^{-7}$~eV.
In addition to the spin-axion searches,
a comagnetometry scheme based on ZULF NMR is capable of measuring the hypothetical spin-dependent gravitational energy of nuclei at the $10^{-17}$~eV level~\cite{RNComagnetometer, RNLiquidstate}.

\section{Conclusion and perspective}

ZULF NMR provides an alternative magnetic resonance modality which is performed in the absence of an applied magnetic field or at extremely small magnetic field.
Over the years, ZULF $\textrm{NMR}$ based on atomic magnetometers has gradually become mature and served as a complementary technique to conventional high-field $\textrm{NMR}$.
This new NMR modality is particularly attractive because of inexpensive, portability, and high absolute magnetic field homogeneity.
In addition to the advantages mentioned above,
ZULF $\textrm{NMR}$ opens the door to applications in,
such as imaging and spectroscopy inside of metal objects or in metallic materials,
which are inaccessible to traditional high-field $\textrm{NMR}$.
In this review, we review the recent advances of ZULF NMR demonstrated in spectroscopy, quantum control, imaging, NMR-based quantum sensing, and tests of fundamental physics.

ZULF NMR is still a relatively young field and its current applications are mainly in laboratories.
The applications of ZULF NMR in industry are still lacking, for example, in the characterization of oil in well logging and food safety testing.
In future, to become a more powerful and versatile technique in research and industry,
a plenty of efforts on theories and experiments are still necessary.
Firstly, the detection sensitivity of ZULF $\textrm{NMR}$ is still lower than that of the state-of-the-art high-field $\textrm{NMR}$ spectrometers.
This can be overcome by several techniques as follows.
The current atomic magnetometers are still far from reaching the quantum-noise-limited sensitivity~\cite{budker2007optical};
several orders of magnitude improvement are possible in the sensitivity,
such as optimizing the magnetic field sensitivity of atomic magnetometers or atomic magnetic gradiometer to subfemtotesla~\cite{kominis2003subfemtotesla, dang2010ultrahigh}, or even attotesla level.
In addition to the atomic magnetometer sensitivity,
it is also possible to improve the nuclear spin polarization through the techniques of parahydrogen-induced polarization (PHIP)~\cite{adams2009reversible, theis2011parahydrogen} and dynamic nuclear polarization ($\textrm{DNP}$)~\cite{barskiy2019zero}.
Secondly, the size of ZULF NMR spectrometer is usually at the level of one meter,
which is mainly limited by lasers, magnetic shield, etc.
Combining with micro-nanotechnology of atomic devices~\cite{shah2007subpicotesla} or anisotropic magnetoresistive sensor~\cite{verpillat2008remote} and microfluidic chip~\cite{ledbetter2008zero}, it is promising to realize a much smaller and lighter ``NMR-on-a-chip" devices,
which could be well suited for real-time chemical analysis in situ without the need to deliver sample to the laboratory.
With the development of nanoscale magnetic sensors, ZULF NMR will promisingly reach single-molecule level~\cite{budker2019extreme}, which is useful for metabolomics at a single-cell level.
Thirdly, the quantum control techniques for ZULF NMR should be developed to analyse the exceedingly complex spectra of small molecules and even macromolecules.
For example, multidimensional ZULF NMR techniques are at the early stage and could be developed as a powerful diagnostic tool for analysing molecular structure.

We believe that the ZULF NMR will progress dramatically over the coming years and shed new lights in applications ranging from materials science, imaging, and quantum information processing to fundamental physics.

~\

\noindent
\medskip
\textbf{Acknowledgements} \par 
We thank Dmitry Budker for valuable discussions. This work was supported by National Key Research and Development Program of China (Grant No. 2018YFA0306600), National Natural Science Foundation of China (Grants Nos. 11661161018, 11927811), Anhui Initiative in Quantum Information Technologies (Grant No. AHY050000),
and USTC Research Funds of the Double First-Class Initiative (Grant No. YD3540002002).

~\

\noindent
\medskip
\textbf{Keywords} \par 
zero- to ultralow-field NMR; atomic magnetometer; spectroscopy; quantum control; imaging; NMR-based quantum devices; tests of fundamental physics.

~\

\noindent
\medskip
\textbf{Conflict of interest} \par 
The authors declare that they have no conflict of interest.


\bibliographystyle{unsrt}
\bibliography{reviews}

\end{document}